\documentclass[10pt,journal,twocolumn]{IEEEtran}
\ifCLASSINFOpdf
   \usepackage[pdftex]{graphicx}
   \DeclareGraphicsExtensions{.pdf,.jpeg,.png}
\else
   \usepackage[dvips]{graphicx}
   \DeclareGraphicsExtensions{.eps}
\fi
\usepackage{amsmath}
\usepackage{amssymb}
\usepackage{amsthm}
\usepackage{cancel}
\usepackage{breqn}
\usepackage{multirow}
\usepackage{rotating}
\usepackage{verbatim}
\usepackage{cool}
\usepackage[caption=false, font=footnotesize]{subfig}

\newcommand{\Expect}{{\rm I\kern-.5em E}}
\DeclareMathAlphabet{\mathitsf}{\encodingdefault}{\sfdefault}{m}{sl}
\DeclareMathAlphabet{\mathitbfsf}{\encodingdefault}{\sfdefault}{bx}{sl}
\newcommand*\dif{\mathop{}\!\mathrm{d}}
\usepackage{cool}
\newtheorem{theorem}{Theorem}[section]
\newtheorem{corollary}{Corollary}[theorem]

\newcommand{\multiline}[2][c]{\begin{tabular}[#1]{@{}c@{}}#2\end{tabular}}
\usepackage[backend=biber,style=ieee,sorting=ynt,doi=false,isbn=false,url=false]{biblatex}
\newcounter{MYtempeqncnt}

\begin{document}

\title{Capacity Analysis of Index Modulations over Spatial, Polarization and Frequency Dimensions}

\author{Pol~Henarejos,%
        ~Ana~I.~P\'{e}rez-Neira, \IEEEmembership{Senior,~IEEE}
\thanks{P. Henarejos is with the Communications Systems Division of Catalonian Technological Telecommunications Center, Barcelona, e-mail: pol.henarejos@cttc.es.}
\thanks{A. P\'{e}rez-Neira is with the Signal and Communication Theory Department of Universitat Polit\`{e}cnica de Catalunya and with the Catalonian Technological Telecommunications Center, e-mail: ana.isabel.perez@upc.edu}%
}

\maketitle

\begin{abstract}
Determining the capacity of a modulation scheme is a fundamental topic of interest. Index Modulations (IM), such as Spatial Modulation (SMod), Polarized Modulation (PMod) or Frequency Index Modulation (FMod), are widely studied in the literature. However, finding a closed-form analytical expression for their capacity still remains an open topic. In this paper, we formulate closed-form expressions for the instantaneous capacity of IM, together with its $2$nd and $4$th order approximations. We show that, in average, the $2$nd approximation error tends to zero for low Signal to Noise Ratio (SNR) and is $o\left(\textrm{SNR}\right)$. Also, a detailed analysis of the ergodic capacity over Rayleigh, Rice and Nakagami-$m$ channel distributions is provided. As application of the capacity analysis, we leverage the proposed expressions to compute the ergodic capacities of SMod for different antenna configuration and correlations, PMod for different channel components and conditions, and FMod for different frequency separations.
\end{abstract}


\IEEEpeerreviewmaketitle

\section{Introduction}
Recent developments in novel transmission schemes present revolutionary mechanisms aimed at increasing the channel capacity. In particular, the use of Multiple-Input Multiple-Output (MIMO) architectures enables a large number of new challenging schemes. For example, one of the implementations of MIMO is currently being deployed by utilising several antennas distributed in a finite region of space. Alternative schemes that do not need channel state information (CSI) at the transmitter end are Vertical Bell Laboratories Layered Space-Time (V-BLAST) \cite{Wolniansky1998,Golden1999}, Spatial Modulation (SMod) \cite{KilfoylePreisigBaggeroer2003,Mesleh2008,Jeganathan2008,Yang2012} or Orthogonal Space-Time Block Codes (OSTBC) \cite{Pascual-Iserte2004,Jorswieck2006}.

Although V-BLAST offers higher capacity, it requires high power budget. On the contrary, OSTBC exploits the full channel diversity at the expense of sacrificing multiplexing gain. Note also that OSTBC can only provide full diversity with $2$ transmitting antennas. SMod strikes a balance between V-BLAST and OSTBC by increasing the channel capacity that OSTBC offers with less power requirements and is a trade-off between V-BLAST and OSTBC. In addition, it also provides more flexibility than OSTBC in the achievable rate that is obtained by increasing the number of transmitting antennas.

SMod can be classified in to the general modulation strategy of Index Modulations (IM), term employed by \cite{Basar2015,Basar2016}. Among a set of possible channels, IM is based on the principle of selecting one specific channel depending on the sequence of bits to be transmitted. Hence, in IM information is conveyed both through electromagnetic radiation and also by identifying the channel that is used for the radiated symbol. For instance, SMod uses different spatially uncorrelated antennas to convey information. Part of the sequence of information bits selects which set of antenna is used for transmission. The other part is used to modulate the electromagnetic wave. The receiver can extract the information from the radiated symbol as well as detecting which channel is being used.

Note that IM is an extension of Shift Keying (SK) scheme. With SK, information is located only in the shifts between different elements of a particular dimension. For instance, Spatial Shift Keying (SSK) conveys the information only by selecting the transmitter antenna \cite{Jeganathan2009}, and it can be interpreted as a particular case of SMod that uses the same radiation pattern for all shifts. Additionally, in multiuser scenarios, SSK and reconfigurable antennas can be used to select which user is intended, pointing a beam to it \cite{Bouida2015,Bouida2016}. SMod uses the same SK principle, but the radiation pattern is modulated according to additional information. Although SMod and SK only transmit through a single channel for a time instant, this can be generalized to an arbitrary number of streams, the so-called Generalized SMod (GSMod). Hence, by activating different channels simultaneously, different streams can be multiplexed and the data rate is also increased. This is well explained in \cite{Ibrahim2016}. Additionally, there are studies about the achievable rate for GSMod \cite{Datta2013} and for Generalized Index Modulation \cite{Datta2016}. These generalized modulations are attractive from the computational complexity point of view and achievable rate. However, the capacity analysis is still open.

In our work, we focus on the capacity of IM systems where transmission is done through a single channel. Authors of \cite{Basnayaka2016} and \cite{Narasimhan2016} also analyse the mutual information of SMod but the expressions are integral-based. To solve that, \cite{Faregh2016} presents a first approximation of the integral-based expression using the Meijer G function. However, this approximation is only valid for multiple-input single-output (MISO) systems.

Exploiting the spatial domain is not the only possible dimension. Polarization or frequency domains can also be used. In the case of polarization, Polarized Modulation (PMod) shifts between different orthogonal polarizations following a sequence of bits in order to radiate a different symbol in each shift \cite{Henarejos2015a,Henarejos2015}. In this case, Polarization Shift Keying (PolSK) is a particular case of PMod that uses the same radiation pattern for all shifts \cite{Benedetto1992}. Finally, for the frequency domain, analogous consideration can be pursued \cite{El-Mahdy2002,Basar2016}. In this case, the transmitter alternates different carriers, where an information symbol is radiated in the selected carrier frequency. Whereas Frequency Shift Keying (FSK) only conveys information in the selected frequency hop, Frequency Index Modulation (FMod) conveys information in the frequency hop as well as in the radiated symbol. In \cite{Datta2016}, a design of a system where spatial and frequency domains coexist jointly is introduced and the achievable rate as the number of maximum bits that this technique can transmit, regardless the channel capacity, is presented.

Although the domains of the previous schemes are different (frequency, space, polarization, etc.), they all share the same principle. Providing closed-form analytical expressions of fundamental metrics of IM (such as capacity) is a paramount task. The first work on channel capacity for the previous schemes is introduced in \cite{Song2004}. Later works such as \cite{Yang2008a,Yang2008} introduce the capacity expression in its integral-based expression and formulate the instantaneous capacity for a single receiver antenna. However, the manipulation of the integral-based expression is rather complicated. In \cite{Rajashekar2014}, the authors extend the previous work to an arbitrary dimension at the receiver. In this case, although the capacity is also expressed in its integral-based expression, upper and lower bounds based on the Jensen inequality are introduced. Finally, in \cite{ZhangYangHanzo2015} another integral-based expression of the capacity is provided.

In \cite{Mesleh2017,Younis2017}, a novel approach entitled Quadrature Spatial Modulation (QSM) is proposed where the spatial bits are modulated in the channel vectors and real and imaginary parts are conveyed separately in different channels. In this approach, the symbols' distribution is optimized accordingly to the channel distribution and thus, the capacity that is obtained is equivalent to the ergodic capacity. However, in contrast to \cite{Mesleh2017}, we consider the channel distribution as a deterministic random variable and we continue developing the expressions provided in \cite{Yang2008a,Yang2008,Rajashekar2014}. Moreover, the approach in \cite{Mesleh2017,Younis2017} requires Channel Distribution Information (CDI) at transmitter side, in order to optimize the distribution of symbols depending on the distribution of the channel. In our approach, the transmitter does not require this information at transmitter side. 

Although IMs are designed to be used without the presence of CSI at transmitter, works such as \cite{Rajashekar2013,Rajashekar2015,Sun2017} exploit this aspect by selecting a subset of channel vectors that maximize the capacity. Note that the proposed approach in this paper can also be used with this kind of schemes, changing dynamically the channel subset affects to the ergodic capacity. Other schemes use CSI at transmitter by designing the optimal precoder \cite{Lee2015}. The consequence of using this approach is expressed in terms of an increase of the SNR at receiver.

The present manuscript provides a closed-form analytic expression for an accurate approximation of IM capacity. In addition, as we do not constrain the channel to a particular distribution, we obtain the expression for a generalized fading channel. Later, we compute the ergodic capacity based on different channel distributions. Finally, to illustrate the usefulness of the proposed formulas, we perform the capacity analysis of IM applied to three physical domains: spatial, polarization and frequency. Although the analysed expression is the same in all the results, its consequences are different depending on the physical domain. A summary of our contributions is listed below:
\begin{itemize}
\item We provide a closed-form analytical expression of the IM capacity. This expression is based on the expansion of the Taylor Series for the expected value of the received signal.
\item We formulate $2$nd and $4$th order approximations and we show that these approximations are accurate.
\item We provide a detailed analysis of the average error incurred by the $2$nd approximation. In addition, we show that the expectation of this error tends to zero when the SNR tends to zero. We also show that this error is an $o\left(\textrm{SNR}\right)$.
\item Since the channel capacity formulation has been established over generalized fading channels, we compute the ergodic capacity for different distributions. In particular, we describe the ergodic capacity of IM over Rayleigh, Rice and Nakagami-$m$ channels. 
\item We analyse IM over three domains: spatial, polarization and frequency using realistic channel models.
\end{itemize}

The remainder of this paper is structured as follows: section \ref{sect:system_model} describes the system model and proposes the capacity of IM for a generalized channel realization. Section \ref{sect:ergodic_capacity} introduces the analytical expressions of the ergodic capacity for three channel distributions, namely Rayleigh, Rice and Nakagami-$m$. Section \ref{sect:remainder} analyses the error of the tight approximations of $2$nd and $4$th order. Analytical results and capacity analysis of different applications of IM are described in detail in section \ref{sect:results}. Finally, section \ref{sect:conclusions} discusses the main conclusions of our work.

\section{System Model and Capacity}
\label{sect:system_model}

Given a discrete time instant, IM over an arbitrary MIMO channel realization with $t$ inputs and $r$ outputs, is defined as
\begin{equation}
\mathbf{y}=\sqrt{\gamma}\mathbf{H}\mathbf{x}(l,s)+\mathbf{n}
\label{eq:sysmod}
\end{equation}
where $\mathbf{y}\in\mathbb{C}^{r}$ is the received vector, $\gamma$ is the averaged SNR, $\mathbf{H}=\left[\mathbf{h}_1\,\ldots\,\mathbf{h}_{t}\right]\in\mathbb{C}^{r\times t}$ is the channel matrix, $\mathbf{x}(l,s)\in\mathbb{C}^{t}$ is the transmitted vector, whose components are expressed as $\mathbf{x}(l,s)|_{[l']}=s\delta(l'-l)$, where $\delta(l)$ is the Dirac delta, $l\in\left[1,t\right]$ is the hopping index, $s\in\mathbb{C}$ is the complex symbol from the constellation $\mathcal{S}$. The AWGN noise is modeled as vector $\mathbf{n}\in\mathbb{C}^{r}\sim\mathcal{CN}\left(\mathbf{0},\mathbf{I}\right)$. In other words, $\mathbf{x}(l,s)$ has only one component different from zero (component $l$) and its value is $s$; thus, the radiated symbol hops among the different channels. Note that MIMO channel can be obtained by simultaneous channels in any of the dimensions.

In this section we do not analyse the statistics of $\mathbf{H}$ yet, as we are only interested in the instantaneous capacity given a channel realisation. $\mathbf{H}$ models effects and specific impairments of the employed domain (spatial, polarization, frequency, etc.). For instance in the case of SMod, $\mathbf{H}$ models the channel and the antennas imperfections; in the case of PMod, $\mathbf{H}$ includes all imperfections of polarization dipoles. Note that the aspects of using reconfigurable antennas can also be reflected in $\mathbf{H}$. In following sections we study the capacity for the different channel statistics.

Since the transmitted vector is determined by $(l,s)$, it is possible to rewrite \eqref{eq:sysmod} as
\begin{equation}
\mathbf{y}=\sqrt{\gamma}\mathbf{h}_ls+\mathbf{n}.
\end{equation}

Thus, the symbol $s$ as well as the hopping index $l$ transmit information, the capacity can be expressed as
\begin{equation}
C=\max\limits_{f_{\mathitsf{S}}(s),f_{\mathitsf{L}}(l)} I(\mathbf{y};s,l)\ [\textrm{bpcu}]
\end{equation}
where $f_{\mathitsf{S}}(s)$ and $f_{\mathitsf{L}}(l)$ are the probability density functions (PDF) of the random variables (RV) of the complex symbol $s$ and hopping index $l$, respectively, and $I(\mathitsf{X},\mathitsf{Y})$ is the mutual information (MI) between RV $\mathitsf{X}$ and $\mathitsf{Y}$. Applying the chain rule \cite{Cover2012}, the MI can be decomposed as
\begin{equation}
I(\mathbf{y};s,l)=I(\mathbf{y};s|l)+I(\mathbf{y};l)\triangleq I_1+I_2
\label{eq:I}
\end{equation}
where $I(\mathbf{y};s|l)$ is the MI between the received vector $\mathbf{y}$ and the transmitted symbol $s$ conditioned to the hopping index $l$, and $I(\mathbf{y};l)$ is the MI between the received vector $\mathbf{y}$ and the hopping index $l$. The $l$ index is formed by a uniform RV in the set $\left[1,t\right]$.

As described in \cite{Cover2012}, $I_1$ is maximized when $s$ presents zero mean complex Gaussian distribution. Thus, for a fixed $l$ index, we obtain
\begin{equation}
I_1=I(\mathbf{y};s|l)=\frac{1}{t}\sum_{l=1}^{t}\log_2\left(\sigma_l^2\right)
\label{eq:ysl}
\end{equation}
where $\sigma_l^2=1+\gamma\|\mathbf{h}_l\|^2$. Note that $\sigma_l^2$ depends on which channel is selected by the $l$ index.

Using the sufficient statistics transformation $\mathitsf{Y}=\frac{\mathbf{h}_l^H}{\|\mathbf{h}\|}\mathitbfsf{Y}$, the second term of \eqref{eq:I} can therefore be expressed as 
\begin{align}
I(\mathbf{y};l)&\equiv I(y;l)\nonumber\\
&=-\mathitsf{H}\left(\mathitsf{Y}|\mathitsf{L}\right)+\mathitsf{H}\left(\mathitsf{Y}\right),
\end{align}
where $y=\sqrt{\gamma}\|\mathbf{h}_l\|s+n$, $n\sim\mathcal{CN}(0,1)$. Therefore $\mathitsf{Y}$ is a RV that, for a given $l$, $\mathitsf{Y}\sim\mathcal{CN}\left(0,\sigma_l^2\right)$.

Hence,
\begin{align}
&I_2=I(\mathbf{y};l)\nonumber\\
&=-\sum_{l=1}^{t}\mathitsf{H}\left(\mathitsf{Y}|\mathitsf{L}=l\right)p_{\mathitsf{L}}(l)-\int_{\mathcal{Y}}f_{\mathitsf{Y}}\left(y\right)\log_2\left(f_{\mathitsf{Y}}\left(y\right)\right)\dif y
\label{eq:Ice}\\
&=-\frac{1}{t}\sum_{l=1}^{t}\log_2\left(\pi e\sigma_l^2\right)\nonumber\\
&\,-\frac{1}{t}\sum_{l=1}^{t}\int_{\mathcal{Y}}f_{\mathitsf{Y}|\mathitsf{L}}\left(y|l\right)\log_2\left(\frac{1}{t}\sum_{l'=1}^{t}f_{\mathitsf{Y}|\mathitsf{L}'}\left(y|l'\right)\right)\dif y
\label{eq:Il2}
\end{align}
where $\mathcal{Y}$ is the domain of $y$.

In \eqref{eq:Il2}, the conditioned pdf $f_{\mathitsf{Y}|\mathitsf{L}}\left(y|l\right)$ is described by the pdf of the zero mean valued complex Gaussian distribution, which is expressed as
\begin{equation}
f_{\mathitsf{Y}|\mathitsf{L}}\left(y|l\right)=\frac{1}{\pi\sigma_l^2}\,e^{-\frac{|y|^2}{\sigma_l^2}}.
\label{eq:fyl}
\end{equation}

The integral-based term \eqref{eq:Il2} can be expressed as
\begin{align}
&\int_{\mathcal{Y}}f_{\mathitsf{Y}|\mathitsf{L}}\left(y|l\right)\log_2\left(\frac{1}{t}\sum_{l'=1}^{t}f_{\mathitsf{Y}|\mathitsf{L}'}\left(y|l'\right)\right)\dif y\\
&=-\log_2\left(t\right)+\Expect_{\mathitsf{Y}|\mathitsf{L}}\left\{\log_2\left(\sum_{l'=1}^{t}\frac{1}{\pi\sigma_{l'}^2}\,e^{-\frac{|y|^2}{\sigma_{l'}^2}}\right)\right\}.
\label{eq:e1}
\end{align}

In order to evaluate \eqref{eq:e1}, we decompose the expectation function into its multivariate Taylor series expansion near the mean \cite{FinneyThomasWeir1992}. Note that whereas \cite{Ibrahim2016} expands the logarithm function, we are decomposing the expectation function. Given a sufficiently differentiable function $g$, the Taylor series for the $g(\mathbf{x})$ function with $\mathbf{x}=(x_1,\ldots,x_N)$ in the proximity of $\mathbf{a}=(a_1,\ldots,a_N)$ is described by the multi-index notation \cite{SaintRaymond1991} as\footnote{$\sum_{|\alpha|=n}$. This corresponds to the sum of all possible combinations such that $|\alpha|=n$. For example, for $N=3$, $\sum_{|\alpha|=2}\mathbf{x}^{\alpha}=x_1x_2+x_1x_3+x_2x_3+x_1^2+x_2^2+x_3^2.$}
\begin{equation}
T\left(g,\mathbf{x},\mathbf{a}\right)=\sum_{n=0}^{\infty}\sum_{|\alpha|=n}\frac{1}{\alpha!}\partial^{\alpha}g\left(\mathbf{a}\right)\left(\mathbf{x}-\mathbf{a}\right)^\alpha,
\end{equation}
where $|\alpha|=\alpha_1+\ldots+\alpha_N$, $\alpha!=\alpha_1!\ldots\alpha_N!$, $\mathbf{x}^{\alpha}=x_1^{\alpha_1}\ldots x_N^{\alpha_N}$ and $\partial^{\alpha}g=\partial^{\alpha_1}\ldots \partial^{\alpha_N}=\frac{\partial^{|\alpha|}g}{\partial x_1^{\alpha_1}\ldots \partial x_N^{\alpha_N}}$.

Thus, given a RV $\mathitbfsf{X}$ with finite moments and such that all of its components are uncorrelated, i.e. $\Expect\left\{\mathitsf{X}_i\mathitsf{X}_j\right\}=\delta_{ij}$, the expectation of $g\left(\mathitbfsf{X}\right)$ can be expressed as
\begin{equation}
\Expect_{\mathitbfsf{X}}\left\{T\left(g,\mathitbfsf{X},\mathbf{a}\right)\right\}=\sum_{n=0}^{\infty}\frac{1}{n!}\sum_{m=1}^N\pderiv[n]{g}{x_m}\left(\mathbf{a}\right)\Expect\left\{\left(\mathitsf{X}_m-a_m\right)^n\right\}.
\label{eq:taymul}
\end{equation}

By considering the Taylor series expansion near the expected value of $\mathitbfsf{X}$, $\mathbf{a}=\mu_{\mathitbfsf{X}}$, then \eqref{eq:taymul} becomes
\begin{equation}
\Expect_{\mathitbfsf{X}}\left\{T\left(g,\mathitbfsf{X},\mu_{\mathitbfsf{X}}\right)\right\}=\sum_{n=0}^{\infty}\frac{1}{n!}\sum_{m=1}^N\pderiv[n]{g}{x_m}\left(\mu_{\mathitbfsf{X}}\right)\vartheta_{\mathitsf{X}_m}^n,
\label{eq:tayex}
\end{equation}
where $\vartheta_{\mathitsf{X}_m}^n$ is the centred $n$th moment of $\mathitsf{X}_m$.

The $n$th moment can be computed by deriving $n$ times the Moment-generating function $M_{\mathitbfsf{X}}\left(t\right)$ \cite{GrinsteadSnell2012} and equating it to zero, which for the multivariate normal case is defined as
\begin{equation}
M_{\mathitsf{X}_i}(t)=e^{\mu_{\mathitsf{X}_i}t+\frac{1}{2}\sigma_{\mathitsf{X}_i}^2t^2},
\end{equation}
where $\mu_{\mathitsf{X}_i}$ and $\sigma_{\mathitsf{X}_i}^2$ are the mean and variance of $\mathitsf{X}_i$, respectively.

Since the received signal is a complex normal RV such that $y=\left(\Re(y),\Im(y)\right)=(y_1,y_2)$, the mean and variance of real and imaginary parts are defined by
\begin{align}
\mu_{y_i}&=\Expect\{y_i\}=0,\,i=1,2\\
\sigma_{y_i}^2&=\Expect\left\{\left(y_i-\Expect\{y_i\}\right)^2\right\}=\frac{\sigma_l^2}{2},\,i=1,2,
\end{align}
where the variance of the real and imaginary parts is the half of the variance of the received signal constraint to $l$, $\sigma_l^2$.
Therefore, the $n$th moment of $y$ is expressed as
\begin{equation}
\vartheta_{y_i}^n=\begin{cases}(n-1)!!\frac{\sigma_l^n}{2^{\frac{n}{2}}} &\text{if \textit{n} is even}\\0&\text{if \textit{n} is odd}\end{cases}
\label{eq:normmoments}
\end{equation}
where $n!!=n(n-2)(n-4)...1$. Assuming that $g(y)$ is symmetric in its derivatives, $\pderiv[n]{g}{y_1}(a)=\pderiv[n]{g}{y_2}(a)$, then \eqref{eq:tayex} can be reduced to
\begin{align}
&\Expect_{\mathitsf{Y}}\left\{T\left(g,y,0\right)\right\}\\
&=g(0)+\sum_{n=1}^{\infty}\frac{\sigma_l^{2n}}{2^{2n-1}n!}\frac{\partial^{2n}g}{\partial y_1^{2n}}(0).
\label{eq:expectay}
\end{align}

Assuming that
\begin{align}
g\left(y\right)&=\log_2\left(\sum_{l'=1}^{t}\frac{1}{\pi\sigma_{l'}^2}\,e^{-\frac{y_1^2+y_2^2}{\sigma_{l'}^2}}\right),
\label{eq:defg}
\end{align}
the first term $g(0)$ is expressed as
\begin{equation}
g(0)=\log_2\left(\sum_{l=1}^{t}\frac{1}{\pi\sigma_l^2}\right).
\end{equation}

Therefore \eqref{eq:e1} is described as
\begin{align}
&\int_{\mathcal{Y}}f_{\mathitsf{Y}|\mathitsf{L}}\left(y|l\right)\log_2\left(\frac{1}{t}\sum_{l'=1}^{t}f_{\mathitsf{Y}|\mathitsf{L}}\left(y|l'\right)\right)\dif y\nonumber\\
&=-\log_2\left(t\right)+\log_2\left(\sum_{l'=1}^{t}\frac{1}{\pi\sigma_{l'}^2}\right)+\sum_{n=1}^{\infty}\frac{\sigma_l^{2n}}{2^{2n-1}n!}\frac{\partial^{2n}g}{\partial y_1^{2n}}(0).
\label{eq:ec1}
\end{align}

Combining \eqref{eq:ec1}, \eqref{eq:Il2} can be expressed as
\begin{align}
&I_2=I(\mathbf{y};l)\nonumber\\
&=-\frac{1}{t}\sum_{l=1}^{t}\log_2\left(\pi e\sigma_l^2\right)+\log_2\left(t\right)-\log_2\left(\sum_{l=1}^{t}\frac{1}{\pi\sigma_l^2}\right)\nonumber\\
&-\frac{1}{t}\sum_{n=1}^{\infty}\frac{1}{2^{2n-1}n!}\frac{\partial^{2n}g}{\partial y_1^{2n}}(0)\sum_{l=1}^t\sigma_l^{2n}.
\label{eq:iyl}
\end{align}

Finally, combining \eqref{eq:ysl} and \eqref{eq:iyl}, we can describe the capacity of IM as
\begin{align}
C&=\log_2\left(e^{-1}H\left(\boldsymbol{\sigma}^2\right)\right)-\sum_{n=1}^{\infty}\frac{A\left(\boldsymbol{\sigma}^{2n}\right)}{2^{2n-1}n!}\frac{\partial^{2n}g}{\partial y_1^{2n}}(0),
\label{eq:finalC}
\end{align}
where $\boldsymbol{\sigma}^n=\left(\sigma_1^n,\ldots,\sigma_t^n\right)^T$, and $A(\cdot)$ and $H(\cdot)$ are the arithmetic and harmonic mean operators, respectively. See Annex \ref{sect:apDeriv24} for second and fourth derivative expressions.

In order to get some insight into \eqref{eq:finalC}, we consider its second and fourth order approximations. Taking the second order, $n=1$, the second derivative of $g\left(y\right)$ at zero is expressed as
\begin{align}
\pderiv[2]{g}{y_1}\left(0\right)&=-\frac{2}{\log(2)}\frac{H\left(\boldsymbol{\sigma}^{2}\right)}{H\left(\boldsymbol{\sigma}^{4}\right)}.
\label{eq:d2r}
\end{align}

Hence, we can obtain a $2$nd order approximation of \eqref{eq:finalC} by
\begin{align}
C&\simeq\log_2\left(H\left(\boldsymbol{\sigma}^2\right)\right)-\frac{1}{\log(2)}\left(1-A\left(\boldsymbol{\sigma}^{2}\right)\frac{H\left(\boldsymbol{\sigma}^{2}\right)}{H\left(\boldsymbol{\sigma}^{4}\right)}\right).
\label{eq:finalC2nd}
\end{align}

Similarly, by taking the fourth order, $n=2$, the fourth derivative of $g(y)$ at zero can be expressed as
\begin{align}
\pderiv[4]{g}{y_1}\left(0\right)&=\frac{12}{\log(2)}\left(\frac{H\left(\boldsymbol{\sigma}^{2}\right)}{H\left(\boldsymbol{\sigma}^{6}\right)}-\frac{H^2\left(\boldsymbol{\sigma}^{2}\right)}{H^2\left(\boldsymbol{\sigma}^{4}\right)}\right)
\label{eq:d4r}
\end{align}
and therefore a $4$th order approximation of \eqref{eq:finalC} can be obtained by
\begin{align}
&C\simeq\log_2\left(H\left(\boldsymbol{\sigma}^2\right)\right)-\frac{1}{\log(2)}\left(1-A\left(\boldsymbol{\sigma}^{2}\right)\frac{H\left(\boldsymbol{\sigma}^{2}\right)}{H\left(\boldsymbol{\sigma}^{4}\right)}\right.\nonumber\\
&\left.+\frac{3}{4}A\left(\boldsymbol{\sigma}^{4}\right)\left(\frac{H\left(\boldsymbol{\sigma}^{2}\right)}{H\left(\boldsymbol{\sigma}^{6}\right)}-\frac{H^2\left(\boldsymbol{\sigma}^{2}\right)}{H^2\left(\boldsymbol{\sigma}^{4}\right)}\right)\right).
\label{eq:finalC4th}
\end{align}

\subsection{Capacity analysis in high SNR}

Under the assumption of high SNR regime, i.e., $\gamma\rightarrow\infty$, we can use \eqref{eq:finalC} at the limit and write
\begin{equation}
\lim_{\gamma\rightarrow\infty}C_{\textrm{IM}}=\lim_{\gamma\rightarrow\infty}\log_2\left(\gamma\right).
\label{eq:hchmC}
\end{equation}

We compare it with the asymptotic capacity of MIMO scheme without CSI. This is expressed as
\begin{equation}
C_{\textrm{MIMO}}=\log_2\left|\mathbf{I}+\frac{\gamma}{t}\mathbf{H}^H\mathbf{H}\right|=\sum_{n=1}^L\log_2\left(1+\frac{\gamma}{t}\lambda_n\right),
\label{eq:mimoC}
\end{equation}
where $\lambda_n$ is the $n$th non-zero eigenvalue of $\mathbf{H}^H\mathbf{H}$ and $L$ is the rank of $\mathbf{H}$. At high SNR, we can denote
\begin{equation}
\lim_{\gamma\rightarrow\infty}C_{\textrm{MIMO}}=\lim_{\gamma\rightarrow\infty}L\log_2\left(\gamma\right).
\label{eq:hmimoC}
\end{equation}

Comparing \eqref{eq:hchmC} and \eqref{eq:hmimoC} we can conclude that, effectively, any MIMO technique where transmitter uses all channel modes (e.g. V-BLAST), increases capacity by a factor of $L$. By using IM instead, the use of channel matrix is constraint to a single column and, therefore, the capacity formula is not multiplied by $L$, i.e. \eqref{eq:hchmC}.

\section{Remainder Analysis}
\label{sect:remainder}

In this section we analyse the expectation of the remainder of the $2$nd order approximation \eqref{eq:finalC2nd}. From \eqref{eq:tayex}, the Taylor series expansion with the remainder term can be written as follows
\begin{equation}
\Expect_{\mathitbfsf{X}}\left\{T_k\left(g,\mathitbfsf{X},\mu_{\mathitbfsf{X}}\right)\right\}=\sum_{n=0}^{k}\frac{1}{n!}\sum_{m=1}^N\pderiv[n]{g}{x_m}\left(\mu_{\mathitbfsf{X}}\right)\vartheta_{\mathitsf{X}_m}^n+R_k\left(g,\mathitbfsf{X},\xi\right),
\end{equation}
where $k$ is the order of the Taylor series expansion and for some $\xi$ in the segment $[0,\mathitsf{X}]$. Since the third moment is zero, we analyse the remainder of order $k=3$. $R_3\left(g,\mathitbfsf{X},\xi\right)$ is found by truncating the sum of \eqref{eq:expectay} at $4$th order and evaluated at some point $\xi\in[0,y]$. Thus, 
\begin{equation}
\Expect_{\mathitsf{Y}}\left\{R_3(C)\right\}=\frac{A\left(\boldsymbol{\sigma}^{4}\right)}{32}\left(\pderiv[4]{g}{y_1}(\xi)+\pderiv[4]{g}{y_2}(\xi)\right).
\label{eq:remdef}
\end{equation}

Expression \eqref{eq:remdef} depends on $\xi$, which also depends on $y$. Since $y$ depends on the SNR, we perform the analysis for very low and very high SNR. First, we state the following theorem:
\begin{theorem}
	The expectation of the remainder $\Expect_{\mathitsf{Y}}\left\{R_3\left(g,y,\mu_y\right)\right\}$ is $0$ when all $\sigma_l^2$ tend to the same value $S$, regardless the value $S$. Hence,
	\begin{equation}
	\lim_{\boldsymbol{\sigma}^2\rightarrow S\mathbf{1}}\Expect_{\mathitsf{Y}}\left\{R_3\left(g,y,\mu_y\right)\right\}=0.
	\end{equation}
	\label{eq:th1}
\end{theorem}
See Annex \ref{sect:apDeriv24} for the proof. 

Based on the Theorem \eqref{eq:th1} we can formulate the following corollary:
\begin{corollary}
	The expectation of the remainder tends to $0$ for low SNR, i.e., when $\gamma\rightarrow 0$.
	\begin{equation}
	\lim_{\gamma\rightarrow 0}\Expect_{\mathitsf{Y}}\left\{R_3(C)\right\}=0.
	\end{equation}
\end{corollary}
\begin{proof}
	If $\gamma\rightarrow 0$, then $\boldsymbol{\sigma}^2\rightarrow \mathbf{1}$, which is the condition of Theorem \ref{eq:th1} for $S=1$ and therefore this concludes the proof.
\end{proof}

For high values of SNR, i.e., $\gamma\rightarrow\infty$, we can announce the following theorem:
\begin{theorem}
	The expectation of the remainder $\Expect_{\mathitsf{Y}}\left\{R_3(C)\right\}$ is $o(\gamma)$ when $\gamma\rightarrow \infty$
	\begin{equation}
	\Expect_{\mathitsf{Y}}\left\{R_3(C)\right\}=o(\gamma).
	\end{equation}
	\label{eq:th2}
\end{theorem}
See Annex \ref{sect:apDeriv24} for the proof.

Finally, we can state that:
\begin{itemize}
	\item For low SNR, the expected error tends to zero.
	\item For high SNR, the expected error tends to a constant that does not depend on the SNR, but on the channel realization instead.
\end{itemize}

In the next section, the $2$nd order approximation is used to obtain detailed expressions for the ergodic capacity.

\section{Ergodic Capacity}
\label{sect:ergodic_capacity}
In the previous sections, we studied the capacity for an arbitrary realization of the channel matrix. In this section we analyse the ergodic capacity for different channel statistics. The ergodic capacity is defined as
\begin{equation}
\bar{C}=\Expect_{\mathbf{H}}\{C\}
\label{eq:deferg}
\end{equation}
where $\Expect_{\mathbf{H}}$ is the expectation over all channel realizations. For the sake of clarity, hereinafter we omit the sub-index in the expectation operator, referring to the expectation over the channel statistics. For fast fading channels or when interleaving is carried out, ergodic capacity is a useful bound.

Although the equation \eqref{eq:deferg} does not show a closed-form expression, we exploit the property of harmonic mean of two RV, $H\left(\boldsymbol{\sigma}^2\right)=\frac{2\sigma_1^2\sigma_2^2}{\sigma_1^2+\sigma_2^2}$. Combined with \eqref{eq:finalC2nd} and after simplifying, we obtain
\begin{align}
C&=1+\log_2\left(\sigma_1^2\right)+\log_2\left(\sigma_2^2\right)\nonumber\\
&-\log_2\left(\sigma_1^2+\sigma_2^2\right)-\frac{1}{\log(2)}\left(1-\frac{1}{2}\left(\frac{\sigma_1^2}{\sigma_2^2}+\frac{\sigma_2^2}{\sigma_1^2}\right)\right).
\end{align}
Assuming that all $\sigma_l^2$ follow the same distribution with the same parameters, we apply the statistical average of \eqref{eq:deferg} to obtain
\begin{align}
\bar{C}&=1+2\Expect\left\{\log_2\left(\sigma_1^2\right)\right\}\nonumber\\
&-\Expect\left\{\log_2\left(\sigma_1^2+\sigma_2^2\right)\right\}-\frac{1}{\log(2)}\left(1-\Expect\left\{\frac{\sigma_1^2}{\sigma_2^2}\right\}\right). 
\label{eq:cergt2}
\end{align}

After mathematical manipulations, Table \ref{tab:ergcaps} summarizes the ergodic capacities of Nakagami-$m$, Rice and Rayleigh channel distributions, where \begin{equation}
\begin{split}
\beta&=\frac{m}{\gamma\Omega}\\
\Upsilon(r,\beta)&=\sum_{j=1}^{r-1}\sum_{k=1}^j\left(\frac{\left(k-1\right)!}{j!}\left(-\beta\right)^{j-k}\right)-\textrm{E}_s(r,\beta)\\
\mathrm{E}_s(r,\beta)&=e^{\beta}\mathrm{E}_i\left(-\beta\right)\sum_{j=0}^{r-1}\frac{\left(-\beta\right)^j}{j!}\\
\mathrm{E}_i(-x)&=-\Gamma(0,x)
\end{split}
\end{equation}
and $\mathrm{E}_i(-x)$ is the Exponential Integral function for negative argument and $\Gamma(s,x)=\int_x^{\infty}t^{s-1}e^{-t}\dif t$ is the Incomplete Upper Gamma Function. Note that the Rayleigh ergodic capacity can be obtained by particularizing the Nakagami-$m$ distribution with $m=1$ and $\Omega=2\varrho^2$, where $\varrho$ is the standard deviation of the Rayleigh RV. The derivations can be found in the Appendix \ref{sect:ap1}. It is important to remark that channel phase distribution does not affect capacity analysis, since only $\sigma_l^2$ are used in the expressions.

\begin{table}[!ht]
	\centering
	\caption{Ergodic capacity of the Nakagami-$m$, Rice and Rayleigh channels}
	\begin{tabular}{c|c}
		\textbf{Channel distribution} & \textbf{Ergodic capacity}\\\hline
		\textbf{Nakagami}-$m$&$\begin{aligned}\bar{C}&=\frac{1}{\log(2)}\left(2\Upsilon\left(mr,\beta\right)-\Upsilon\left(2mr,2\beta\right)\right.\\&\left.+\left(1+mr\beta^{-1}\right)\beta^{mr}e^{\beta}\Gamma\left(1-mr,\beta\right)-1\right)\end{aligned}$\\ \hline
		\textbf{Rice} & $\begin{aligned}\bar{C}&=\frac{1}{\log(2)}\sum_{k=0}^{\infty}\frac{e^{-\frac{\lambda}{2}}\left(\frac{\lambda}{2}\right)^k}{k!}\left(2\Upsilon\left(r+k,\beta\right)-\Upsilon\left(2r+k,2\beta\right)\right.\\&\left.+\left(1+\frac{r+k}{\beta}\right)\beta^{r+k}e^{\beta}\Gamma\left(1-r-k,\beta\right)-1\right)\end{aligned}$\\ \hline
		\textbf{Rayleigh} & $\begin{aligned}\bar{C}&=\frac{1}{\log(2)}\left(2\Upsilon\left(r,\beta\right)-\Upsilon\left(2r,2\beta\right)\right.\\&\left.+\left(1+r\beta^{-1}\right)\beta^re^{\beta}\Gamma\left(1-r,\beta\right)-1\right)\end{aligned}$\\ \hline
	\end{tabular}
	\label{tab:ergcaps}
\end{table}
\begin{table}[!ht]
\centering
\caption{Antenna Correlation Matrices}
\begin{tabular}{c|c|c|c}
&One antenna & Two antennas & Four antennas\\\hline
$R_{\textrm{TX}}$&$1$ & $\begin{pmatrix}1 & \alpha\\\alpha^* & 1 \end{pmatrix}$ & $\begin{pmatrix}1 & \alpha^{1/9} & \alpha^{4/9} & \alpha \\\alpha^{*1/9} & 1 & \alpha^{1/9} & \alpha^{4/9}\\\alpha^{*4/9} & \alpha^{*1/9} & 1 & \alpha^{1/9}\\\alpha & \alpha^{*4/9} & \alpha^{*1/9} & 1 \end{pmatrix}$\\\hline
$R_{\textrm{RX}}$&$1$ & $\begin{pmatrix}
1 & \beta\\\beta^* & 1\end{pmatrix}$ & $\begin{pmatrix}1 & \beta^{1/9} & \beta^{4/9} & \beta \\\beta^{*1/9} & 1 & \beta^{1/9} & \beta^{4/9}\\\beta^{*4/9} & \beta^{*1/9} & 1 & \beta^{1/9}\\\beta & \beta^{*4/9} & \beta^{*1/9} & 1 \end{pmatrix}$\\\hline
\end{tabular}
\label{tab:spcorr}
\end{table}
\section{Results}
\label{sect:results}

In this section we present some results stemming from the work described in the previous sections. First, we compare the proposed approximations \eqref{eq:finalC2nd} and \eqref{eq:finalC4th}. In order to compute the system capacity we first generate $N$ channel realisations. Then, we compute the instantaneous capacity for each channel realisation following different approaches. Finally, we average the obtained results amongst all realisations. For very large $N$, this procedure is equivalent to calculating the ergodic capacity. Unless explicitly stated otherwise, both the transmitter and the receiver have two inputs and two outputs, $t=r=2$.

Following subsections describe two studies. In the first section, the proposed approximations are validated and compared with the integral-based expressions introduced in \cite{Yang2008} and \cite{Rajashekar2014}. In the section, we employ the proposed approximations to compare and analyse different applications of IM. In more detail, we counterpose IM applied to the frequency, spatial and polarization domains.

\subsection{Analytical Results}
\begin{figure}[!ht]
  \centering
    \includegraphics[width=0.9\linewidth]{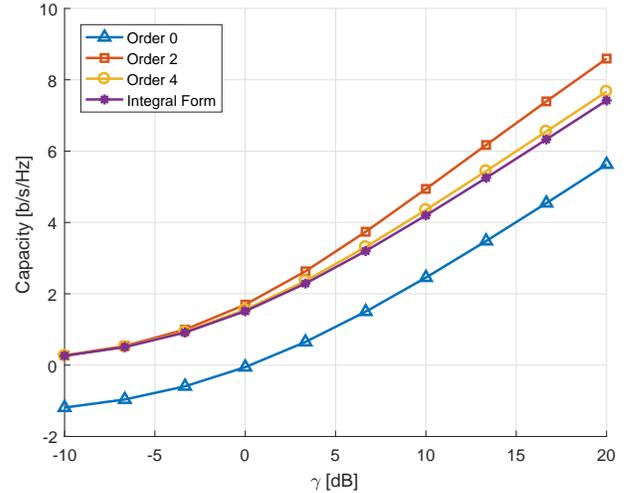}
    \caption{Average instantaneous capacity approximations for different orders compared to the integral-based expression for $t=2$ and $r=2$.}
    \label{fig:PlotCapacityInt_t2r2}
\end{figure}

This section studies analytical results using an arbitrary channel matrix. Fig. \ref{fig:PlotCapacityInt_t2r2} depicts the computed ergodic capacity of the proposed approximations ($2$nd and $4$th orders) and its integral-based expression, described in \cite{Yang2008}. Note that the channel matrix follows the Rayleigh distribution. This figure shows how the approximation of the integral-based expression evolves. This is described by Taylor's Theorem \cite{GrinsteadSnell2012} and is validated in the figure. As the order is increased, the error between the approximation and the integral-based expression decreases notably.

\begin{figure}[!ht]
  \centering
    \includegraphics[width=0.9\linewidth]{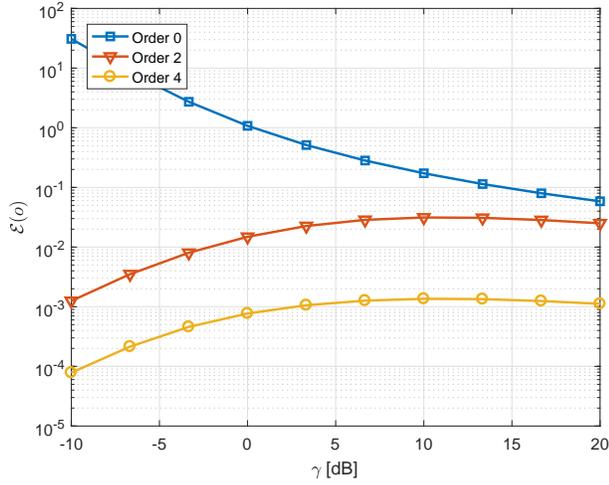}
    \caption{Normalized error of the different approximation orders for $t=2$ and $r=2$.}
    \label{fig:PlotCapacityIntDiff_t2r2}
\end{figure}

Fig. \ref{fig:PlotCapacityIntDiff_t2r2} illustrates the normalised error of each approximations relative to the integral-based expression. Note that the normalised error compares the performance of the different approximation orders and the integral form expression implicitly and is defined as
\begin{equation}
\mathcal{E}(o)=\frac{\left|\sum_nC_{\textrm{Order=o}}(n)-C_{\textrm{Integral}}(n)\right|^2}{\left|\sum_nC_{\textrm{Integral}}(n)\right|^2}.
\end{equation}
With this figure, first we show that the normalised error tends to zero as we increase the order. Second, the figure validates theorems \eqref{eq:th1} and \eqref{eq:th2}, which state that the error tends to zero for low SNR and is constant for high SNR, respectively. 

\begin{figure}[!ht]
	\centering
	\subfloat[$t=1$]{\includegraphics[width=0.49\linewidth,clip=true]{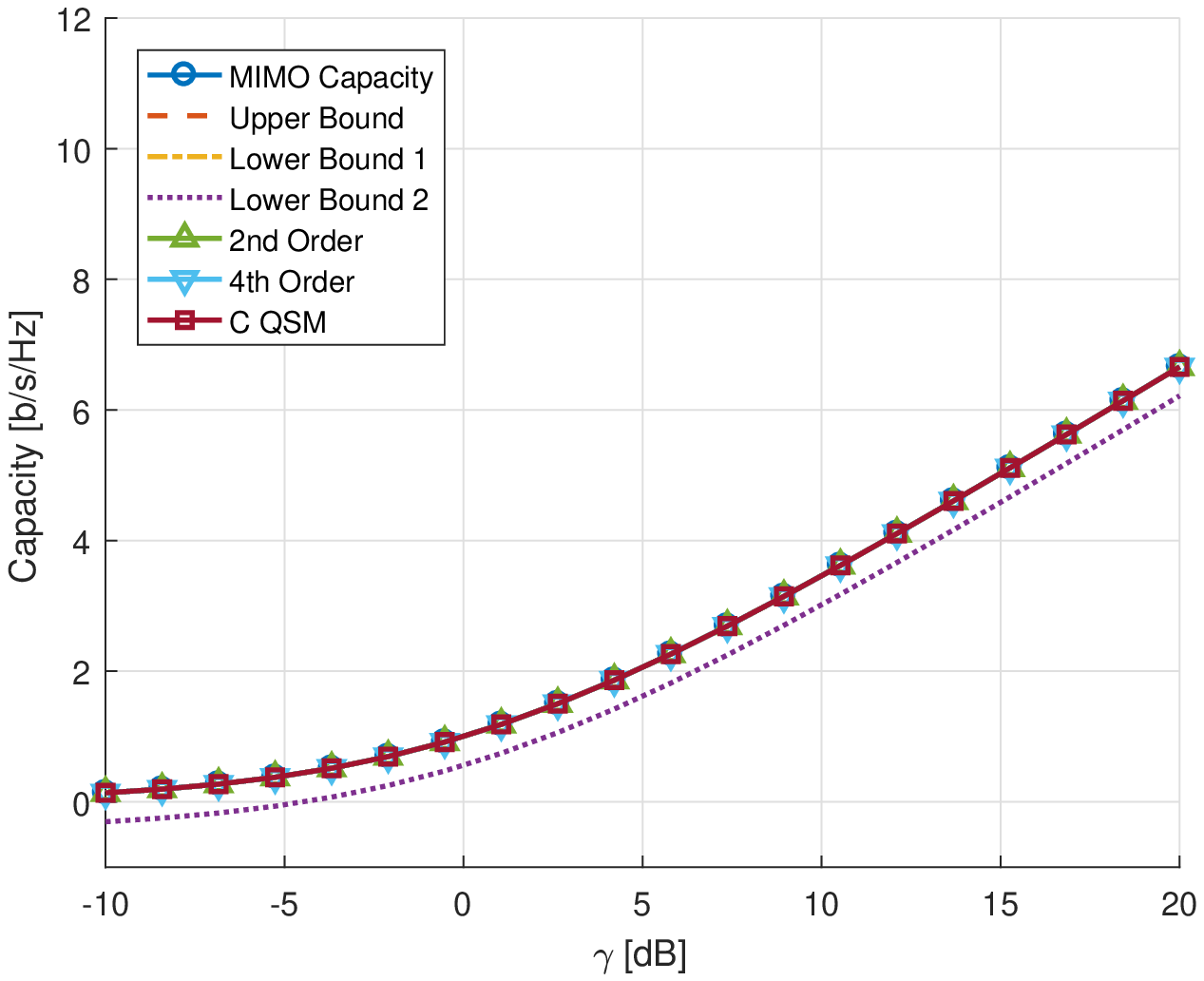}}\hfill
	\subfloat[$t=2$]{\includegraphics[width=0.49\linewidth,clip=true]{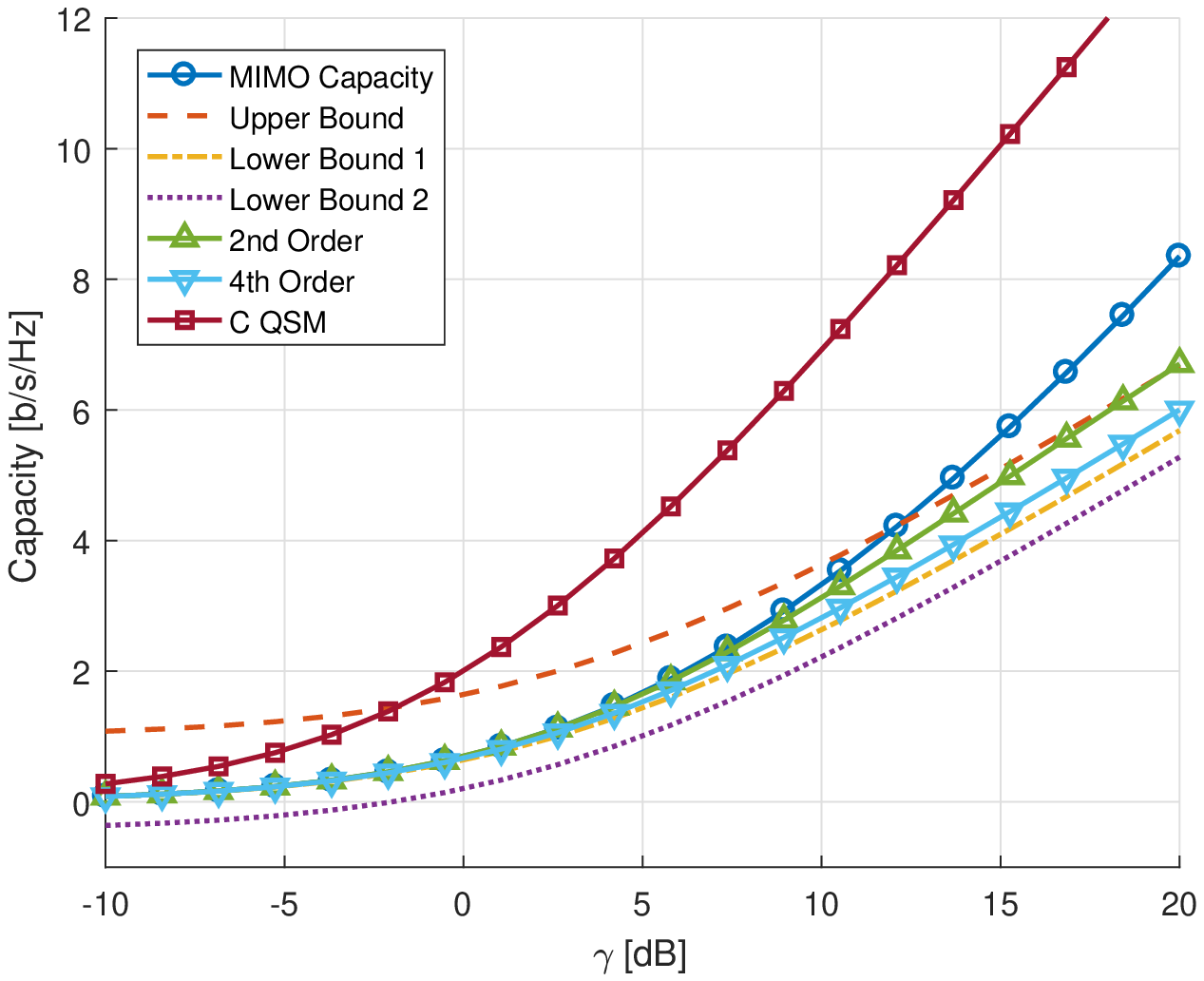}}\hfill
	\subfloat[$t=4$]{\includegraphics[width=0.49\linewidth,clip=true]{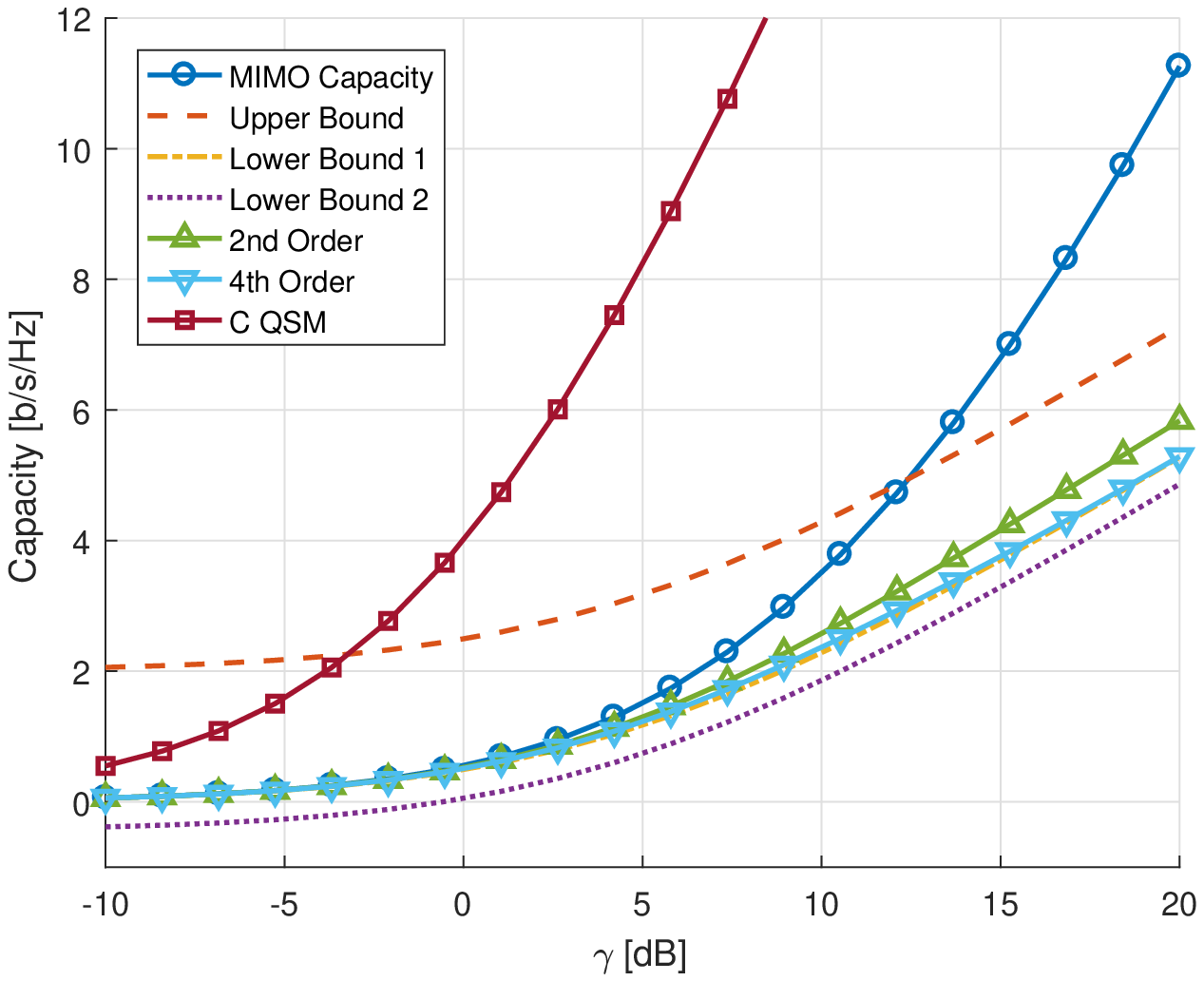}}\hfill
	\subfloat[$t=8$]{\includegraphics[width=0.49\linewidth,clip=true]{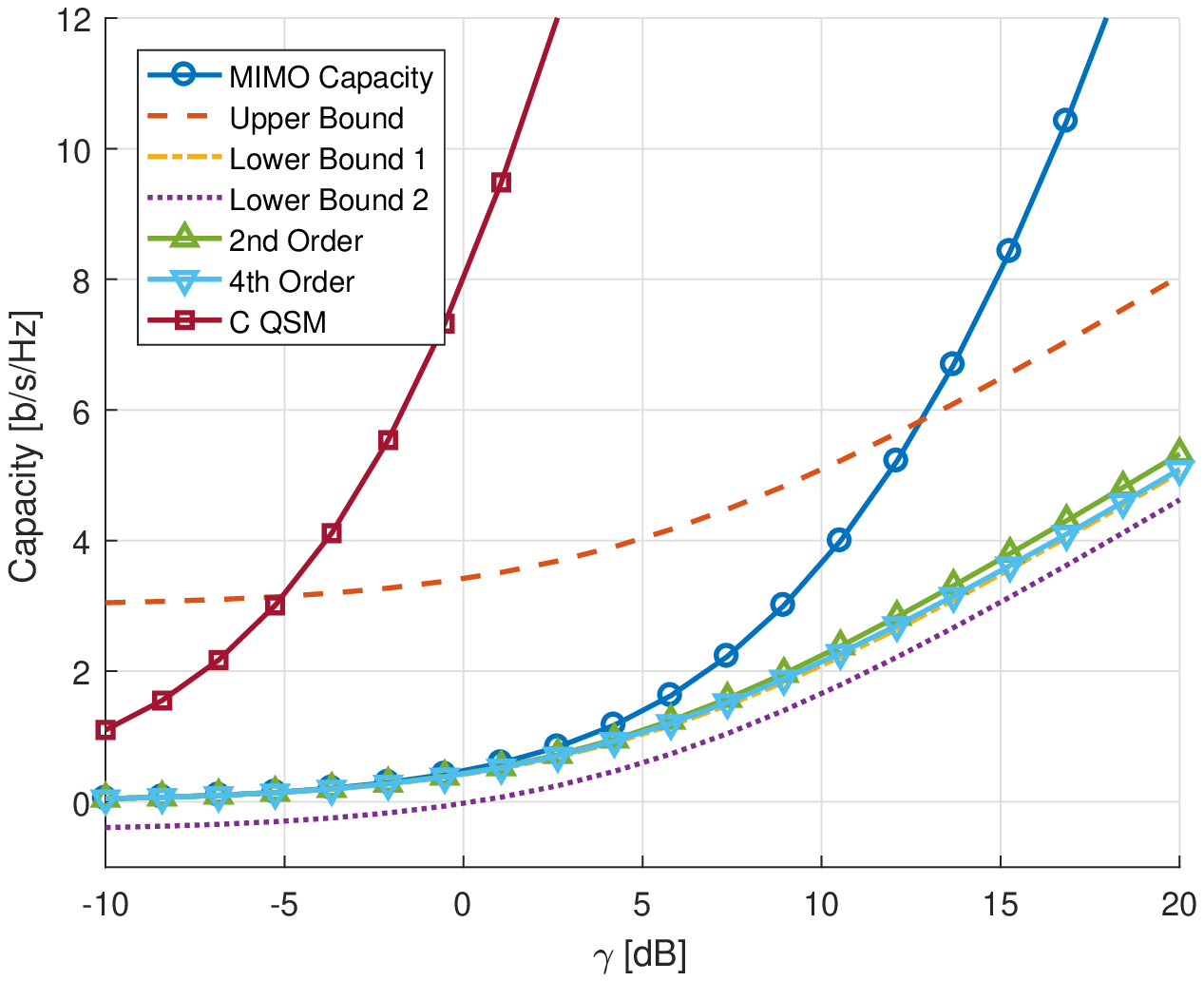}}\hfill
	\caption{Approximations of the capacity for different orders compared to the upper and lower bounds described in \cite{Rajashekar2014} and the capacity of QSM of \cite{Younis2017}.}
	\label{fig:PlotCapBounds_t2r2}
\end{figure}

Fig. \ref{fig:PlotCapBounds_t2r2} shows the second and fourth order approximations and the upper and lower bounds described in \cite{Rajashekar2014}. The figure is obtained by different number of antennas at transmission ($t=1,2,4$ and $8$). With this figure, we demonstrate that the proposed approximations are placed between both bounds. The MIMO capacity in absence of CSIT is also depicted, which is described by \eqref{eq:mimoC} and (9) of \cite{Mesleh2017} (denoted as $C_{\textrm{H Dependent}}$), as well as the capacity of QSM described in \cite{Younis2017} (labelled as \emph{C CQSM}). This figure reflects the effect of the rank of the channel, $L$. For instance, whereas the slope of IM is $2/5$, the slope of MIMO is $4/5$, increased by $L=2$ with respect to IM. Note that for $t=1$, the bounds and the performance of different orders approximations are completely overlapped, since it is equivalent to a SIMO scheme.

Finally, the last analysis is performed in terms of computational complexity. This analysis is particularly interesting because a priori it seems more plausible the use of the integral-based expression since it is more precise. However, the computational complexity is much expensive compared to the proposed approximations. Table \ref{tab:cplx} describes the average computational complexity in terms time consumption and its relative increment with respect to the Order $0$ approximation. It shows clearly that the proposed approximations can reduce the time consumption more than $4$ times with respect to the integral-based expression. Therefore, the proposed approximations represent a fair trade-off between precision and computational complexity. Note that, although the time consumption depends on the implementation and the machine utilized for the computation, this table reflects the notably difference of using instantaneous closed-form expressions and integral-form expressions.

\begin{table}[!ht]
	\centering
	\caption{Computational Complexity}
	\begin{tabular}{c|c|c|c}
		& \textbf{\multiline{Average Time\\Consumption [$\mu s$]}} & \textbf{\multiline{Relative\\increment [$\%$]}} & \textbf{\multiline{Precision\\(MSE) [$10^{-3}$]}}\\\hline
		\textbf{Order $0$} & $27.758$ & $-$ & $26$\\\hline
		\textbf{Order $2$} & $31.466$ & $13.36$ & $5.5$ \\\hline
		\textbf{Order $4$} & $38.802$ & $26.43$ & $0.3$ \\\hline
		\textbf{\multiline{Integral-based\\expression \cite{Yang2008}}} & $151.115$ & $404.62$ & $-$ \\\hline
	\end{tabular}
	\label{tab:cplx}
\end{table}

\subsection{Applications of Index Modulations}
In this section we discuss the applicability of IM to different domains: spatial, polarization and frequency. We note that in all three cases, the capacity in bits per channel use is the one shown. Therefore, the impact of more bandwidth in FMod does not come up in the study that follows.
\subsubsection{Spatial Modulation}
SMod consists in applying IM to the spatial domain. Using several antennas at transmission, the transmitter can modulate additional information deciding which antenna uses for transmission. Assuming the channels are uncorrelated, the receiver can obtain the additional information by detecting which antenna is being used at transmission. This scheme is specially interesting where transmitters are equipped with many antennas, such as in Long Term Evolution (LTE) or Wi-Fi (IEEE 802.11n and upwards). 

To evaluate the capacity of SMod under realistic scenarios we employ the channel model described by 3GPP \cite{3GPPTS36,3GPPTS36a}. The channel profile corresponds to that described by the Extended Typical Urban model (ETU), with independent realizations. This implies that consecutive channel realizations are not correlated and, thus, do not depend on the Doppler frequency shift. Spatial channels are uncorrelated if the separation between antennas is greater than $\lambda/2$, which is desirable when SMod is used. However, due to imperfections of the transmitter and receiver, antennas can be correlated in different levels. We use $2$ levels defined in the specifications: no correlation and high correlation.
Antenna correlation matrices are defined by Table \ref{tab:spcorr} and \ref{tab:spcorrpara}.
\begin{table}[!ht]
\centering
\caption{Antenna Correlation Parameters}
\begin{tabular}{c|c|c}
 & $\alpha$ & $\beta$ \\\hline
No correlation & $0$ & $0$ \\\hline
Medium correlation & $0.3$ & $0.9$ \\\hline
High correlation & $0.9$ & $0.9$ \\\hline
\end{tabular}
\label{tab:spcorrpara}
\end{table}

\begin{figure}[!ht]
  \centering
    \subfloat[Low Correlation]{\includegraphics[width=0.49\linewidth]{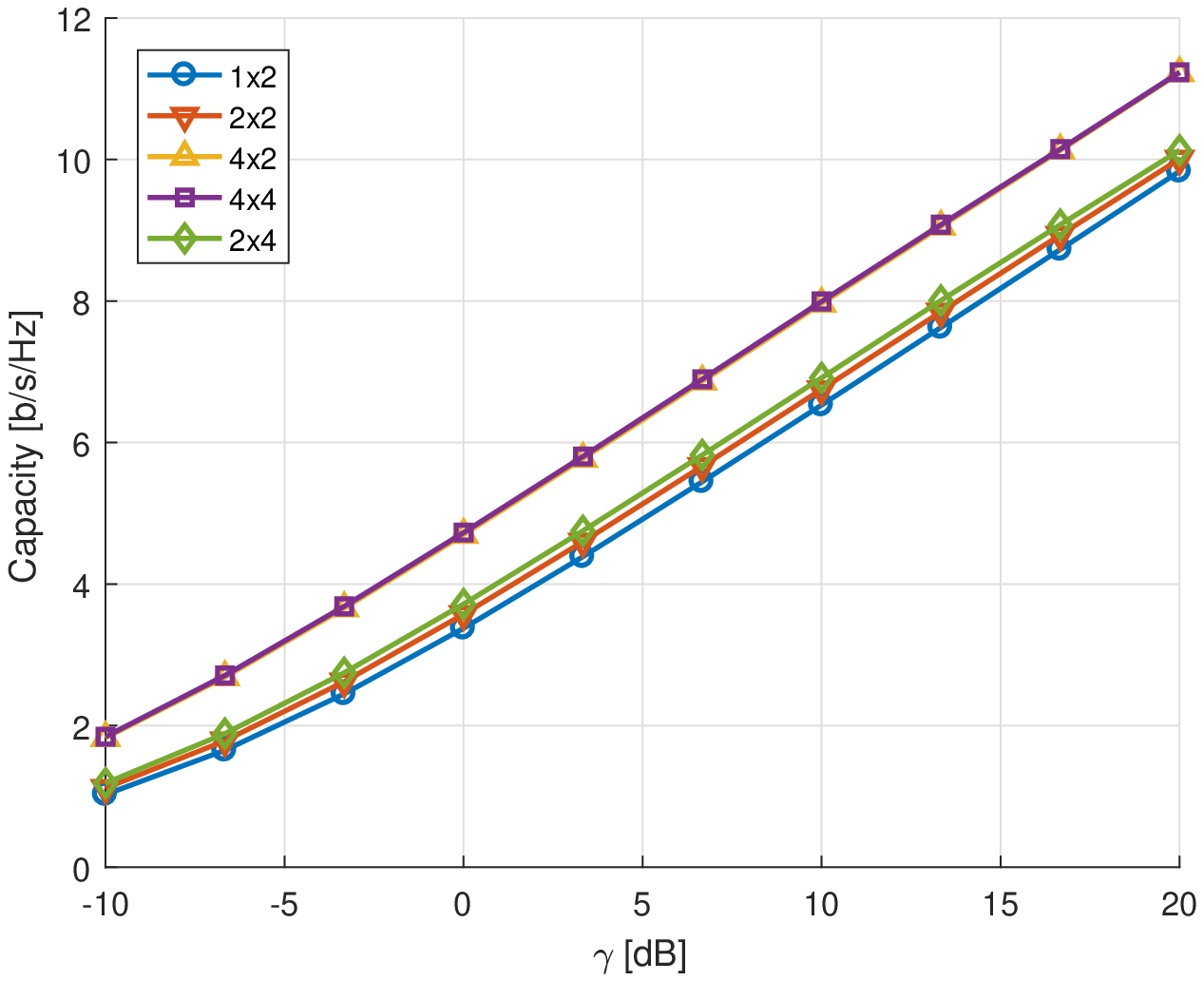}\label{fig:PlotCapScenariosSpatial_Low}}\hfill
    \subfloat[High Correlation]{\includegraphics[width=0.49\linewidth]{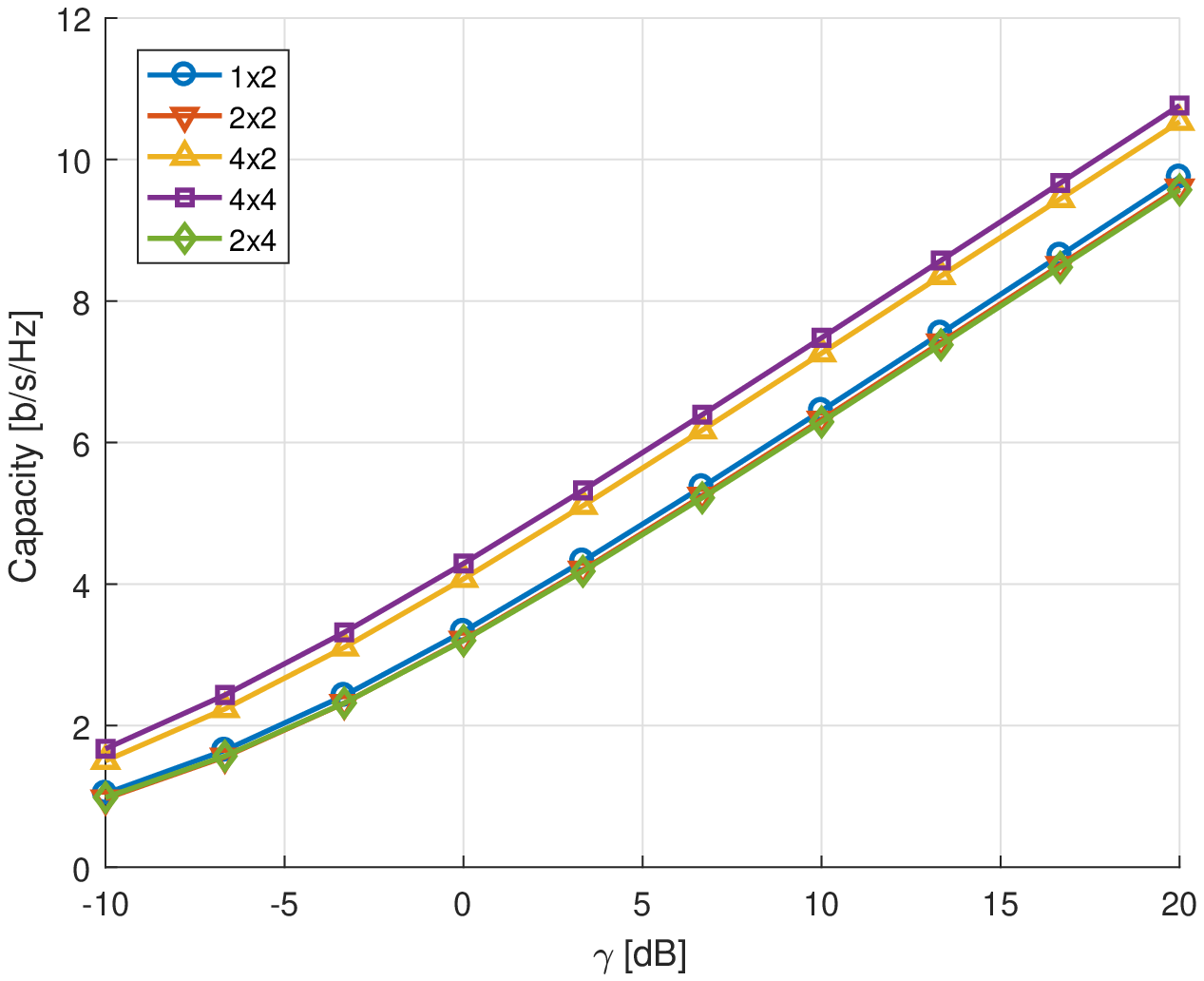}
    \label{fig:PlotCapScenariosSpatial_High}}\hfill
    \caption{Capacity evaluation for different antennas at transmission and reception applying IM to the spatial domain of LTE ETU channel with low and high antenna correlation.}
\end{figure}

Fig. \ref{fig:PlotCapScenariosSpatial_Low} and \ref{fig:PlotCapScenariosSpatial_High} depict the capacity of SMod under ETU channel conditions for no correlation and high correlation of antennas at transmission and reception, and for different number of antennas. From these figures, several appreciations arise:
\begin{enumerate}
	\item Increasing the number of antennas at transmission increases the capacity in SMod. For instance, the highest capacity is achieved for the $4\times 4$ mode. 
	\item As expected, the capacity of SMod decreases when antenna correlation is introduced. SMod exploit spatial diversity inherently by hopping between spatial channels. If antennas are correlated, spatial channels are also correlated, diversity is not fully exploited and the capacity decreases.
	\item The presence of antenna correlation may underperform other modes. For instance $2\times 4$ underperforms $2\times 2$ in the presence of high correlation.
	\item As expected, $1\times 2$ obtains the lower capacity.
\end{enumerate}
\subsubsection{Polarized Modulation}
In contrast to the previous section, polarization domain is not widely used in mobile radio communications. Mobile terminals are handed in different ways with different physical orientations, without respecting the polarization direction. Nevertheless, it is still possible to employ the polarization domain with fixed terminals, such as those generally used in satellite services. Moreover, in satellite communications it is not possible to exploit the spatial diversity due to the correlation between spatial paths. Hence, in these scenarios the polarization dimension takes an important relevance and becomes more challenging.

\begin{figure}[!ht]
  \centering
    \includegraphics[width=0.9\linewidth]{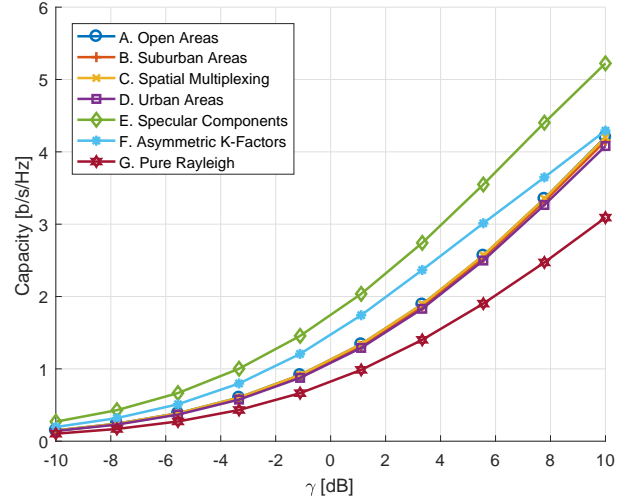}
    \caption{Capacity evaluation for different scenarios applying IM to the polarization domain of Land Mobile Satellite channel.}
    \label{fig:PlotCapScenarios}
\end{figure}

We use the channel model proposed in \cite{Sellathurai2006}, which describes different scenarios for land mobile satellite communications. It incorporates parameters such as correlation between rays, direct, specular, and diffuse rays; as well as cross-talk between inputs and other features. By tuning these parameters, different scenarios such as urban, suburban or maritime environments can be modelled. Fig. \ref{fig:PlotCapScenarios} depicts the IM capacity in different scenarios using the polarization domain. With this figure, we aim to employ the $2$nd order approximation to compare different satellite channels in terms of capacity. Thanks to it, we are able to classify which environmental conditions are more suitable for IM. 

Specifically, we consider PMod activating only vertical or horizontal polarization in each hop. We use typical parameters such as a sampling frequency of $F_s=33.6$ kHz, carrier centred at L-band and a mobility of $5$ m/s. Whilst scenarios such as open areas, suburban areas, spatial multiplexing, urban areas, and Rice channels with asymmetric $K$-factors attain the same capacity, scenarios with specular components increase the capacity by an additional $1$ b/s/Hz with respect to the others. These scenarios achieve a better performance, as specular components can be added to the direct ray. Note that, in general, specular components are present in scenarios where there is a strong reflection, such as, for instance, the maritime scenario, due to the strong reflection of the sea.

\subsubsection{Frequency Index Modulation}
In the frequency dimension, the index modulation is achieved by hopping between available subcarriers. On one hand, flat fading channels imply that all subcarriers are affected by the same channel magnitude and phase and therefore the receiver has to estimate which subcarrier is used by the transmitter. This approach requires high frequency isolation and power budget. These channels are typical in scenarios where there is a strong Line of Sight (LOS). Note that FMod complements FSK, where the information is placed only in the shifts, but differs from Frequency Hopping (FH). In the latter case, no information is placed in the hops and its objective is to exploit frequency diversity and increase security at physical level.

On the other hand, frequency selective channels generate rich frequency diversity since the subcarriers are affected by different channel magnitudes and phases. In this case, the frequency isolation is not critical as previous since the receiver can exploit the CSI to estimate the used subcarrier more accurately. These channels are typical in scenarios with multipath.

\begin{figure}[!ht]
  \centering
    \includegraphics[width=0.9\linewidth]{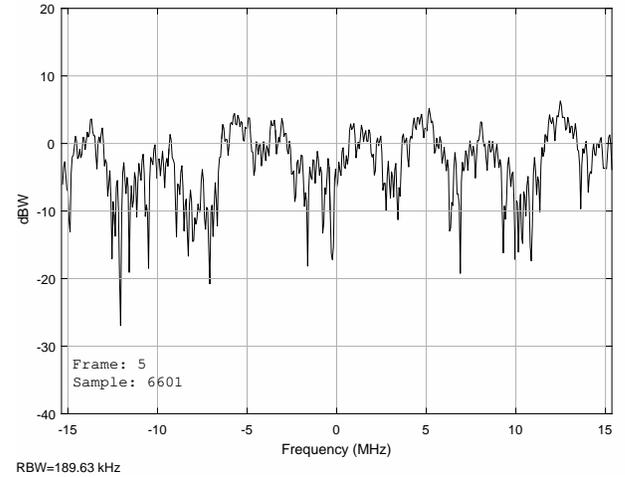}
    \caption{Snapshot of the spectrum of the ETU channel model.}
    \label{fig:PlotSpectrum}
\end{figure}

In order to exploit frequency selective property, frequency hops cannot be adjacent. Intuitively we could think that the more separated subcarriers are, the better capacity the system will achieve. But this is not true. Frequency selective channels present frequency fading randomly at different subcarriers. Fig. \ref{fig:PlotSpectrum} depicts an example of snapshot of ETU channel. It can be appreciated that choosing too separated subcarriers may not be the best strategy.

\begin{figure}[!ht]
  \centering
    \includegraphics[width=0.9\linewidth]{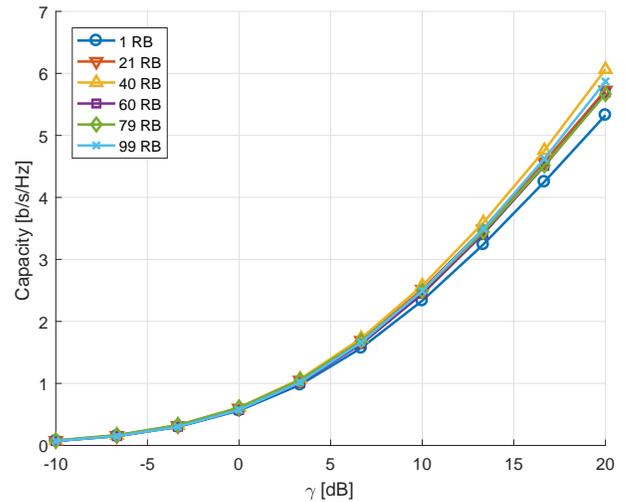}
    \caption{Capacity evaluation for several frequency separations applying IM to the frequency domain of LTE ETU channel.}
    \label{fig:PlotCapScenariosFreq}
\end{figure}

Fig. \ref{fig:PlotCapScenariosFreq} depicts the capacity of FMod for different separations, in Resource Blocks units ($1$ RB = $180$ kHz). Clearly, $1$ RB of separation achieves the lowest capacity. The separation of subcarriers determines the performance in terms of capacity and is specific for each channel profile. Depending on the environment, there are some subcarriers that provide higher gains in contrast to others. For instance, in Fig. \ref{fig:PlotSpectrum}, subcarriers at $5$, $7$ and $12$ MHz provide higher gains compared with the subcarriers at $6$, $10$ or $13$ MHz. Thus, having the maximum separation does not guarantee the maximum capacity.

An additional important aspect is that, in contrast to SMod or PMod, the performance in the capacity is the same for low SNR regime. This means that in low SNR regime, the separation of subcarriers is not relevant and does not affect the performance.  Note that in low SNR, the spectrum can be masked by the noise floor and thus, the predominant term is the noise contribution instead of the subcarrier separation. Also, FMod occupies more bandwidth when number of hops is increased.

\section{Conclusions}
\label{sect:conclusions}
In this paper we present a closed-form expression of the IM capacity, \eqref{eq:finalC}, as well as two closed-forms of its $2$nd and $4$th order approximations, which are \eqref{eq:finalC2nd} and \eqref{eq:finalC4th}, respectively. These expressions are valid for different channel distributions, and provide an approximation to the integral-based expression. We analytically demonstrate that the expectation of the error of the $2$nd and $4$th order approximations tends to zero for low SNR and is $o(\mathrm{SNR})$. This fact is illustrated with several simulations. We also compute the ergodic capacity for Rayleigh, Rice, and Nakagami-$m$ channels based on its $2$nd order approximation, summarized in Table \ref{tab:ergcaps}. These expressions allow to find the ergodic channel capacity without computing the instantaneous capacity over many channel realisations. Finally, we apply the capacity analysis of IM to three physical properties: spatial, polarization and frequency. With SMod, the number of antennas at transmitter and receiver increases the capacity, as well as the correlation between antennas; with PMod, the maximum capacity is achieved when the channel contains specular components; with FMod, the separation between subcarriers affects directly the capacity of the system only in medium and high SNR regimes.

\section*{Acknowledgement}
This work has received funding from the Spanish Ministry of Economy and Competitiveness (Ministerio de Economia y Competitividad) under project TEC2014-59255-C3-1-R and from Satellite Network of Experts (SatNEx) IV of European Space Agency (ESA).

\appendices
\section{Derivations of the ergodic capacity}
\label{sect:ap1}

Based on \eqref{eq:cergt2}, we need to compute three expectations: $E_1=\Expect\left\{\log_2\left(\sigma_1^2\right)\right\}$, $E_2=\Expect\left\{\log_2\left(\sigma_1^2+\sigma_2^2\right)\right\}$ and $E_3=\Expect\left\{\frac{\sigma_1^2}{\sigma_2^2}\right\}$.

The distribution of $\sigma_l^2$ is closely related to the distribution of $h_{il}$. In the following subsections we study the ergodic capacity by using the $2$nd order approximation \eqref{eq:finalC2nd}, particularized by $t=2$, for different common channel distributions, Rice, Nakagami-$m$ and Rayleigh channel distributions, where the latter can be derived from Nakagami-$m$ with $m=1$.

\subsection{Nakagami-$m$ Channel}

In the Nakagami-$m$ channel the envelope of each component of the channel matrix $|h_{ij}|\sim\textrm{Nakagami-}m(\Omega)$ follows the Nakagami-$m$ distribution with a variance of $\Omega$. In this case, $\|\mathbf{h}_l\|^2$ is a sum of $r$ Gamma distributed RV with parameters $\left(m,\Omega/m\right)$. Assuming that each component $h_{ij}$ follows the same distribution parameters, the sum of Gamma distributions with same parameters is also a Gamma distribution, where the shape parameter is the sum of the individual parameters. Thus, the pdf of $\|\mathbf{h}_l\|^2$ is denoted by
\begin{equation}
f_{\textrm{Gamma}\left(mr,\Omega/m\right)}(x)=\frac{m^{mr}}{\Gamma(mr)\Omega^{mr}}x^{mr-1}e^{-m\frac{x}{\Omega}}.
\label{eq:pdfngamma}
\end{equation}

The transformation of $\sigma_l^2$ is also a RV with the following pdf
\begin{equation}
f_{\sigma_l^2}(x)=\frac{m^{mr}}{\Gamma(mr)\left(\gamma\Omega\right)^{mr}}(x-1)^{mr-1}e^{-m\frac{x-1}{\gamma\Omega}}
\end{equation}
and moments $\mu_{\sigma_l^2}=r\gamma\Omega+1$, $\sigma_{\sigma_l^2}^2=r(\gamma\Omega)^2$.

$E_1$ can be solved as follows
\begin{align}
&\Expect\left\{\log_2\left(\sigma_l^2\right)\right\}=\int_0^{\infty}\log_2\left(1+\gamma x\right)f_{\textrm{Gamma}\left(mr,\Omega/m\right)}(x)dx\nonumber\\
&=\frac{1}{\log(2)}\Upsilon\left(mr,\beta\right).
\label{eq:exptotgamma}
\end{align}

In order to compute $\Expect\left\{\log_2\left(\sigma_1^2+\sigma_2^2\right)\right\}$ we exploit the fact that $\log_2\left(\sigma_1^2+\sigma_2^2\right)=1+\log_2\left(1+\frac{\gamma}{2}\left(\|\mathbf{h}_1\|^2+\|\mathbf{h}_2\|^2\right)\right)$. Thus,
\begin{align}
&\Expect\left\{\log_2\left(\sigma_1^2+\sigma_2^2\right)\right\}\nonumber\\
&=1+\int_0^{\infty}\log_2\left(1+\frac{\gamma}{2} x\right)f_{\textrm{Gamma}\left(2mr,\Omega/m\right)}(x)dx\nonumber\\
&=1+\frac{1}{\log(2)}\Upsilon(2mr,2\beta).
\label{eq:expgammasum}
\end{align}

Finally, to compute $E_3$, we recall that $\sigma_l^2$ are independent RV and thus
\begin{equation}
\Expect\left\{\frac{\sigma_1^2}{\sigma_2^2}\right\}=\Expect\left\{\sigma_1^2\right\}\Expect\left\{\frac{1}{\sigma_2^2}\right\}
\end{equation}
where the second expectation is defined as follows
\begin{align}
\Expect\left\{\frac{1}{\sigma_2^2}\right\}&=\int_0^{\infty}\frac{1}{1+\gamma x}f_{\textrm{Gamma}\left(mr,\Omega/m\right)}(x)dx\nonumber\\
&=\beta^{mr}e^{\beta}\Gamma\left(1-mr,\beta\right).
\label{eq:invsigma}
\end{align}
Hence, $E_3$ is solved as
\begin{equation}
\Expect\left\{\frac{\sigma_1^2}{\sigma_2^2}\right\}=\left(1+mr\beta^{-1}\right)\beta^{mr}e^{\beta}\Gamma\left(1-mr,\beta\right).
\label{eq:expgammaratio}
\end{equation}

Finally, we can express the Ergodic Capacity for the Nakagami-$m$ Channel joining \eqref{eq:exptotgamma}, \eqref{eq:expgammasum} and \eqref{eq:expgammaratio} in \eqref{eq:cergt2}.

Note that the Rayleigh channel is obtained when Nakagami-$m$ is particularized for $m=1$ and $\Omega=2\varrho^2$.

\subsection{Rice Channel}

The Rice channel is such that each component $h_{ij}\sim\mathcal{CN}(\nu,\varrho^2)$. In this case, $\|\mathbf{h}_l\|^2$ is a Non-Central Chi-Squared distribution of $2r$ degrees of freedom $\chi_{2r}^2(\lambda)$ whose pdf is denoted by
\begin{equation}
f_{\chi_{2r}^2(\lambda)}(x)=\frac{e^{-\frac{\lambda}{2}}}{2\varrho^2}e^{-\frac{x}{2\varrho^2}}\left(\frac{x}{\lambda\varrho^2}\right)^{\frac{r-1}{2}}I_{r-1}\left(\sqrt{\frac{\lambda x}{\varrho^2}}\right)
\label{eq:pdfnchis}
\end{equation}
where $\lambda=2r\nu^2$ is the non-centrality parameter and $I_{a}(2x)=x^a\sum_{k=0}^{\infty}\frac{x^{2k}}{k!\Gamma\left(a+k+1\right)}$ is the modified Bessel function of first kind of $a$ degrees of freedom.

The transformation of $\sigma_l^2$ is also a RV with the following pdf
\begin{equation}
f_{\sigma_l^2}(x)=\frac{e^{-\frac{\lambda}{2}}}{2\gamma\varrho^2}e^{-\frac{x-1}{2\gamma\varrho^2}}\left(\frac{x-1}{\lambda\gamma\varrho^2}\right)^{\frac{r-1}{2}}I_{r-1}\left(\sqrt{\frac{\lambda(x-1)}{\gamma\varrho^2}}\right)
\end{equation}
and moments $\mu_{\sigma_l^2}=2r\gamma\left(\nu^2+\varrho^2\right)+1$, $\sigma_{\sigma_l^2}^2=4r\gamma^2\varrho^2\left(2\nu^2+\varrho^2\right)$.

In order to compute the expectation of $\log_2\left(\sigma_l^2\right)$, we decompose a Non-Central Chi-Squared RV as a mixture of Central Chi-Squared RV as follows 
\begin{equation}
f_{\chi_{2r}^2(\lambda)}(x)=\sum_{k=0}^{\infty}\frac{e^{-\frac{\lambda}{2}}\left(\frac{\lambda}{2}\right)^k}{k!}f_{\chi_{2r+2k}^2}(x).
\end{equation}
Thus, $E_1$ can be expressed as
\begin{align}
&\Expect\left\{\log_2\left(\sigma_l^2\right)\right\}\nonumber\\
&=\int_0^{\infty}\log_2\left(1+\gamma x\right)\sum_{k=0}^{\infty}\frac{e^{-\frac{\lambda}{2}}\left(\frac{\lambda}{2}\right)^k}{k!}f_{\chi_{2r+2k}^2}(x)dx\nonumber\\
&=\frac{1}{\log(2)}\sum_{k=0}^{\infty}\frac{e^{-\frac{\lambda}{2}}\left(\frac{\lambda}{2}\right)^k}{k!}\Upsilon(r+k,\beta)
\label{eq:expsum}
\end{align}

To find $E_2$, we exploit the result obtained in \eqref{eq:expgammasum}, thus obtaining
\begin{align}
&\Expect\left\{\log_2\left(\sigma_1^2+\sigma_2^2\right)\right\}=1+\int_0^{\infty}\log_2\left(1+\frac{\gamma}{2} x\right)f_{\chi^2_{4r}(\lambda)}(x)dx\nonumber\\
&=1+\frac{1}{\log(2)}\sum_{k=0}^{\infty}\frac{e^{-\frac{\lambda}{2}}\left(\frac{\lambda}{2}\right)^k}{k!}\Upsilon(2r+k,2\beta).
\label{eq:expricesum}
\end{align}

Similarly, we use the result in \eqref{eq:expgammaratio} to compute $E_3$ as
\begin{align}
&\Expect\left\{\frac{\sigma_1^2}{\sigma_2^2}\right\}\\
&=\sum_{k=0}^{\infty}\frac{e^{-\frac{\lambda}{2}}\left(\frac{\lambda}{2}\right)^k}{k!}\left(1+\frac{r+k}{\beta}\right)\beta^{r+k}e^{\beta}\Gamma\left(1-r-k,\beta\right).
\label{eq:expriceratio}
\end{align}

Finally, we can express the Ergodic Capacity for a Rice Channel joining \eqref{eq:expsum}, \eqref{eq:expricesum} and \eqref{eq:expriceratio} in \eqref{eq:cergt2}.

Note that Rayleigh distribution can also be obtained from the Rice distribution imposing $\lambda=0$. In this case, the mixture of Central Chi-Squared RV degenerates into an indetermination for $k=0$. This problem can be solved examining $\lim_{\lambda\rightarrow 0}\bar{C}$, whose solution agrees with the analytical expression in Table \ref{tab:ergcaps}.

\section{Proof of Theorems}
\label{sect:apDeriv24}
In this appendix we describe the second and fourth derivatives of $g(y)$, involved in section \ref{sect:system_model}. We recall that
\begin{equation}
\begin{split}
g\left(y\right)&=\log_2\left(\sum_{l=1}^{t}\frac{1}{\pi\sigma_{l}^2}\,e^{-\frac{y_1^2+y_2^2}{\sigma_{l}^2}}\right)\\
g(y)&\doteq\log_2(f_2(y)),
\end{split}
\end{equation}
where we define $f_n(y)\doteq\sum_{l=1}^{t}\frac{1}{\pi\sigma_{l}^n}\,e^{-\frac{y_1^2+y_2^2}{\sigma_{l}^2}}$.
We first compute the first four derivatives of $g(y)$ as a function of $f_n(y)$, see \eqref{eq:gderiv}.

\begin{figure*}[!t]
	\normalsize
	\setcounter{MYtempeqncnt}{\value{equation}}
	\begin{equation}
	\begin{split}
	\pderiv{g}{y_1}(y)&=\frac{1}{\log 2}\frac{\pderiv{f_2}{y_1}(y)}{f_2(y)}\\
	\pderiv[2]{g}{y_1}(y)&=\frac{1}{\log 2}\left(\frac{\pderiv[2]{f_2}{y_1}(y)}{f_2(y)}-\frac{\left(\pderiv{f_2}{y_1}(y)\right)^2}{f_2^2(y)}\right)\\
	\pderiv[3]{g}{y_1}(y)&=\frac{1}{\log 2}\left(\frac{\pderiv[3]{f_2}{y_1}(y)}{f_2(y)}-3\frac{\pderiv[2]{f_2}{y_1}(y)\pderiv{f_2}{y_1}(y)}{f_2^2(y)}+2\frac{\left(\pderiv{f_2}{y_1}(y)\right)^3}{f_2^3(y)}\right)\\
	\pderiv[4]{g}{y_1}(y)&=\frac{1}{\log 2}\left(\frac{\pderiv[4]{f_2}{y_1}(y)}{f_2(y)}-\frac{4\pderiv[3]{f_2}{y_1}(y)\pderiv{f_2}{y_1}(y)+3\left(\pderiv[2]{f_2}{y_1}(y)\right)^2}{f_2^2(y)}+12\frac{\pderiv[2]{f_2}{y_1}(y)\left(\pderiv{f_2}{y_1}(y)\right)^2}{f_2^3(y)}-6\frac{\left(\pderiv{f_2}{y_1}(y)\right)^4}{f_2^4(y)}\right)
	\end{split}
	\label{eq:gderiv}
	\end{equation}
	\setcounter{MYtempeqncnt}{\value{equation}}
	\setcounter{equation}{\value{MYtempeqncnt}}
	\hrulefill
	\vspace*{4pt}
\end{figure*}

The derivatives of $f_2(y)$ are denoted as follows:
\begin{equation}
\begin{split}
\pderiv{f_2}{y_1}(y)&=-2y_1f_4(y)\\
\pderiv[2]{f_2}{y_1}(y)&=2\left(2y_1^2f_6(y)-f_4(y)\right)\\
\pderiv[3]{f_2}{y_1}(y)&=4y_1\left(3f_6(y)-2y_1^2f_8(y)\right)\\
\pderiv[4]{f_2}{y_1}(y)&=4\left(3f_6(y)-12y_1^2f_8(y)+4y_1^4f_{10}(y)\right)
\end{split}
\label{eq:f2deriv}
\end{equation}

Thus, combining \eqref{eq:gderiv} and \eqref{eq:f2deriv} and after some mathematical arrangements, we can simplify the derivatives, described in \eqref{eq:gderivf2}.

\begin{figure*}[!t]
	\normalsize
	\setcounter{MYtempeqncnt}{\value{equation}}
	\begin{equation}
	\begin{split}
	\pderiv{g}{y_1}(y)&=-\frac{2y_1}{\log 2}\frac{f_4(y)}{f_2(y)}\\
	\pderiv[2]{g}{y_1}(y)&=\frac{2}{\log 2}\left(\frac{2y_1^2f_6(y)-f_4(y)}{f_2(y)}-\frac{2y_1^2f_4^2(y)}{f_2^2(y)}\right)\\
	\pderiv[3]{g}{y_1}(y)&=\frac{4y_1}{\log 2}\left(\frac{3f_6(y)-2y_1^2f_8(y)}{f_2(y)}+3f_4(y)\frac{2y_1^2f_6(y)-f_4(y)}{f_2^2(y)}-4y_1^2\frac{f_4^3(y)}{f_2^3(y)}\right)\\
	\pderiv[4]{g}{y_1}(y)&=\frac{4}{\log 2}\left(\frac{3f_6(y)-12y_1^2f_8(y)+4y_1^4f_{10}(y)}{f_2(y)}+\frac{8y_1^2f_4(y)\left(3f_6(y)-2y_1^2f_8(y)\right)-3\left(2y_1^2f_6(y)-f_4(y)\right)^2}{f_2^2(y)}\right.\\
	&\left.+24y_1^2f_4^2(y)\frac{2y_1^2f_6(y)-f_4(y)	}{f_2^3(y)}-24y_1^4\frac{f_4^4(y)}{f_2^4(y)}\right)
	\end{split}
	\label{eq:gderivf2}
	\end{equation}
	\setcounter{MYtempeqncnt}{\value{equation}}
	\setcounter{equation}{\value{MYtempeqncnt}}
	\hrulefill
	\vspace*{4pt}
\end{figure*}

To prove the theorem \ref{eq:th1} we compute the limit of the fourth derivative of $g(y)$. 
\begin{proof}
	The equality $H\left(\boldsymbol{\sigma}^n\right)=A\left(\boldsymbol{\sigma}^n\right)$ holds when $\sigma_1^n=\ldots=\sigma_t^n=S$. Thus, we compute the limit of the fourth derivative assuming this equality. Computing the limit is equivalent to compute the limit of each $f_n$ term of the fourth derivative\footnote{We denote $\boldsymbol{\sigma}^2\rightarrow S\mathbf{1}$ is equivalent to $\sigma_l^2\rightarrow S, \forall l\in[1,t]$.}. In any case, it is straightforward that
	\begin{equation}
	\lim_{\boldsymbol{\sigma}^2\rightarrow S\mathbf{1}}f_n(y)=\frac{t}{\pi S^{\frac{n}{2}}}e^{-\frac{y_1^2+y_2^2}{S}}=S^{-\frac{n}{2}}F(y).
	\end{equation} 
	Hence, we have that
	\begin{equation}
	\lim_{\boldsymbol{\sigma}^2\rightarrow S\mathbf{1}}f_n^p(y)f_m^q(y)=S^{-\frac{pn+qm}{2}}F^{p+q}(y).
	\label{eq:limSF}
	\end{equation}
	After few mathematical manipulations using \eqref{eq:limSF}, we can ensure that
	\begin{equation}
	\lim_{\boldsymbol{\sigma}^2\rightarrow S\mathbf{1}}\pderiv[4]{g}{y_1}(y)=0.
	\end{equation}
\end{proof}

To prove the theorem \ref{eq:th2} we proceed as previously but in the limit with $\gamma\rightarrow\infty$. 
\begin{proof}
	It can be seen that $\lim_{\gamma\rightarrow\infty}\pderiv[4]{g}{y_1}(\xi)=0$ and $\lim_{\gamma\rightarrow\infty}A\left(\boldsymbol{\sigma}^n\right)=\gamma^{\frac{n}{2}}A\left(\boldsymbol{\kappa}^n\right)$, where $\boldsymbol{\kappa}^n=\left(\|\mathbf{h}_1\|^n\ldots\|\mathbf{h}_t\|^n\right)^T$. Hence, the remainder takes the form of the indetermination $\infty\cdot 0$. To solve it we first compute the following limit:
	\begin{equation}
	\lim_{\gamma\rightarrow\infty}f_n(\xi)=\gamma^{-\frac{n}{2}}\sum_{l=1}^t\frac{1}{\pi\|\mathbf{h}_l\|^n}e^{-\frac{\xi_1^2+\xi_2^2}{\sigma_l^2}}=\gamma^{-\frac{n}{2}}\frac{t}{\pi}H\left(\boldsymbol{\kappa}^n\right)^{-1}
	\end{equation}
	Hence, we have that
	\begin{equation}
	\begin{split}
	\lim_{\gamma\rightarrow\infty}A\left(\boldsymbol{\sigma}^k\right)f_n^p(\xi)&=\gamma^{\frac{k-np}{2}}A\left(\boldsymbol{\kappa}^n\right)\left(\frac{t}{\pi}\right)^{p}H\left(\boldsymbol{\kappa}^n\right)^{-p}\\
	&=\left\{\begin{array}{ll}0 & \textrm{if}\ k<pn\\A\left(\boldsymbol{\kappa}^k\right)\left(\frac{t}{\pi}\right)^{p}H\left(\boldsymbol{\kappa}^n\right)^{-p} & \textrm{if}\ k=pn\\\infty & \textrm{if}\ k>pn\end{array}\right..
	\end{split}
	\label{eq:limAFP}
	\end{equation}
	Applying \eqref{eq:limAFP} to the fourth remainder, it is reduced to
	\begin{equation}
	\begin{split}
	&\lim_{\gamma\rightarrow\infty}\frac{A\left(\boldsymbol{\sigma}^4\right)}{32}\left(\pderiv[4]{g}{y_1}(\xi)+\pderiv[4]{g}{y_2}(\xi)\right)\\
	&=\lim_{\gamma\rightarrow\infty}\frac{3}{4\log 2}A\left(\boldsymbol{\sigma}^4\right)\left(\frac{f_6(\xi)}{f_2(\xi)}-\frac{f_4^2(\xi)}{f_2^2(\xi)}\right)\\
	&=\frac{3}{4\log 2}A\left(\boldsymbol{\kappa}^4\right)\left(\frac{H\left(\boldsymbol{\kappa}^{2}\right)}{H\left(\boldsymbol{\kappa}^{6}\right)}-\frac{H\left(\boldsymbol{\kappa}^{2}\right)^2}{H\left(\boldsymbol{\kappa}^{4}\right)^2}\right).
	\end{split}
	\label{eq:limR3}
	\end{equation}
	Hence, the expectation of the remainder is a constant that does not depend on $\gamma$, regardless the $\xi$ value. We can conclude that $\Expect\left\{R_3\left(g,y,\mu_y\right)\right\}=o(\gamma)$ since
	\begin{equation}
	\lim_{\gamma\rightarrow\infty}\frac{\Expect\left\{R_3\left(g,y,\mu_y\right)\right\}}{\gamma}=0,
	\end{equation}
	which is the definition of the $o(\gamma)$.
	
	Note that a function $f(x)$ can be asymptotically constant, despite of $x$. In this case, $f(x)$ is an $o(x)$, since $f(x)$ can be always upperbounded by $x$.
\end{proof} 

Note that \eqref{eq:limR3} also holds to the theorem \eqref{eq:th1} when $\boldsymbol{\sigma}^2\rightarrow S\mathbf{1}$, which in this case the limit is $0$.

\newpage

\printbibliography
\begin{IEEEbiography}[{\includegraphics[width=1in,height=1.25in,clip,keepaspectratio]{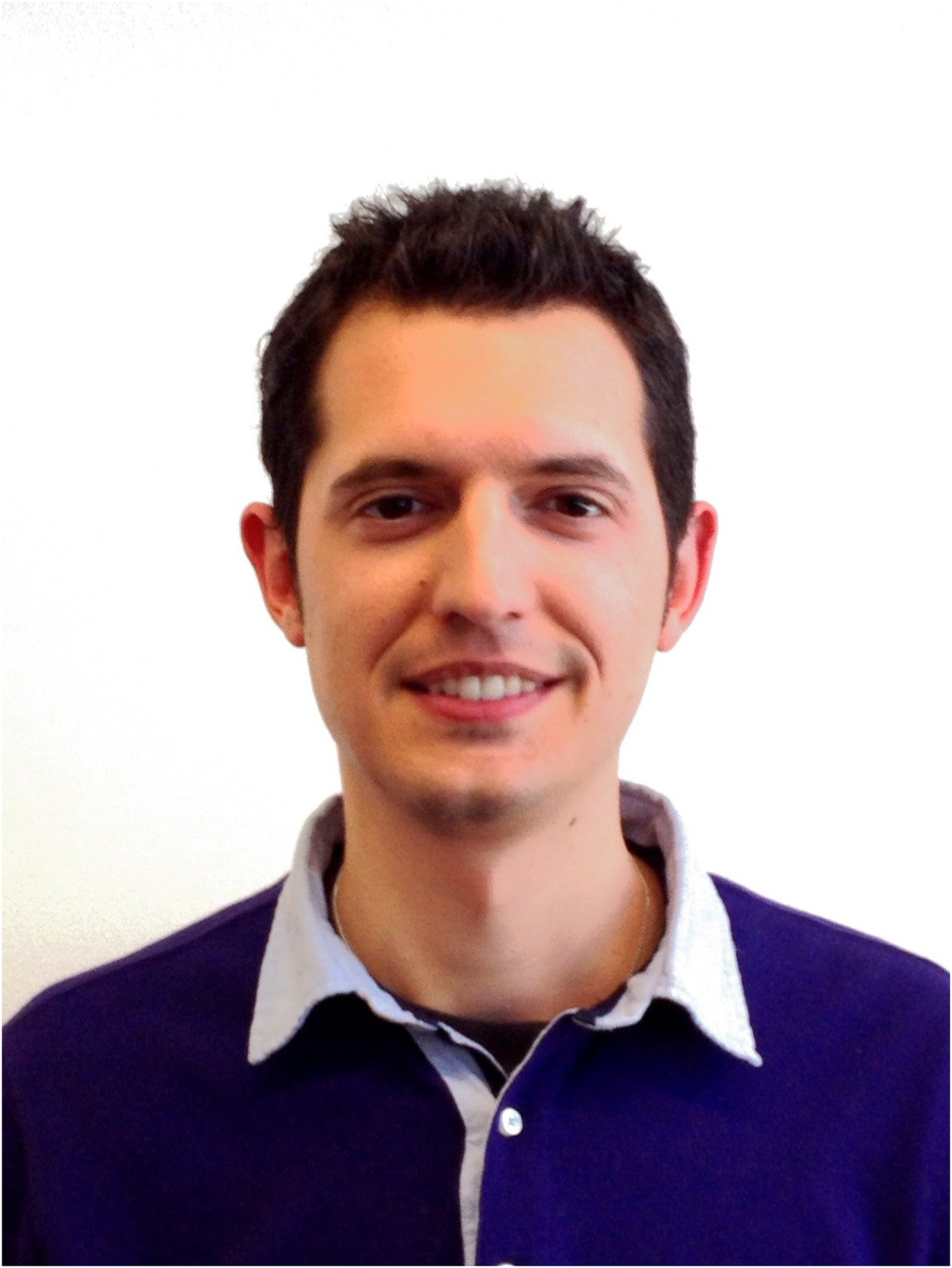}}]{Pol Henarejos}
 received the Telecommunication Engineering degree from the Telecommunication Engineering High School from Barcelona (ETSETB) of Technical University of Catalonia (UPC) in May 2009. In 2012, he obtained the Master of Science in Research on Information and Communication Technologies from UPC. In 2017, he obtained the Ph.D. degree from UPC. He joined the CTTC in January 2010 in Engineering area and he worked prototyping the physical layer communication technologies using software development. He also obtained a knowledge in LTE technologies thanks to industrial contracts for implementing the physical layer of LTE, BGAN and Li-Fi standards. He participated in European projects such as PHYDYAS and FANTASTIC-5G and with industrial contracts. Additionally, he participated in projects funded by the European Space Agency. He is also the promoter of the CASTLE Platform. Currently he is researching on new technologies based on multimedia satellite communications with dual polarization, implementation of physical stack of many standards and enabling 5G technologies on flexible multicast and broadcast communications.
\end{IEEEbiography}

\begin{IEEEbiography}[{\includegraphics[width=1in,height=1.25in,clip,keepaspectratio]{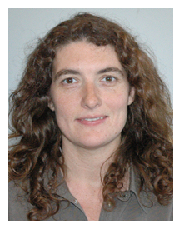}}]{Ana I. P\'{e}rez-Neira}
	is full professor at UPC (Technical University of Catalonia) in the Signal Theory and Communication department. Her research topic is signal processing for communications and currently she is working in multi-antenna and multicarrier signal processing, both, for satellite communications and wireless systems. She has been in the board of directors of ETSETB (Telecom Barcelona) from 2000-03 and Vicerector for Research at UPC (2010- 13). She created UPC Doctoral School (2011). Currently, she is Scientific Coordinator at CTTC (Centre Tecnolgic de Teleco- municacions de Catalunya). Since 2008 she is member of EURASIP BoD (European Signal Processing Association) and since 2010 of IEEE SPTM (Signal Processing Theory and Methods). She is the coordinator of the European project SANSA and of the Network of Excellence on satellite communications, financed by the European Space Agency: SatnexIV. She has been the leader of 20 projects and has participated in over 50 (10 for European Space Agency). She is author of 50 journal papers (20 related with Satcom) and more than 200 conference papers (20 invited). She is co-author of 4 books and 5 patents (1 on satcom). She has been guest editor in 5 special issues and currently she is editor of IEEE Transactions on Signal Processing and of Eurasip Signal Processing and Advances in Signal Processing. She has been the general chairman of IWCLD09, EUSIPCO11, EW14 and IWSCS14. She has participated in the organization of ESA conference 1996, SAM04 and she is the general chair of next ASMS16. She is the chair of next IEEE ICASSP20.

\end{IEEEbiography}

\end{document}